\documentclass[%
 reprint,
 superscriptaddress,
 amsmath,amssymb,
 aps,
]{revtex4-2}

\usepackage{graphicx}
\usepackage{dcolumn}
\usepackage{bm}
\usepackage{xcolor}
\usepackage{subcaption}

\usepackage[colorlinks=true, linkcolor=blue, citecolor=blue, urlcolor=blue]{hyperref}
\usepackage{comment}

\begin{document}


\title{Simulation of dark photon sensitivity in $\eta \rightarrow \gamma e^+e^-$ at HIAF}    

\author{Zaiba Mushtaq}
\affiliation{Institute of Modern Physics, Chinese Academy of Sciences, Lanzhou 730000, China}
\affiliation{School of Nuclear Science and Technology, University of Chinese Academy of Sciences, Beijing 100049, China}

\author{Rong Wang}
\email{Corresponding author: rwang@impcas.ac.cn}
\affiliation{Institute of Modern Physics, Chinese Academy of Sciences, Lanzhou 730000, China}
\affiliation{School of Nuclear Science and Technology, University of Chinese Academy of Sciences, Beijing 100049, China}

\author{Xiaolin Kang}
\email{Corresponding author: kangxiaolin@cug.edu.cn}
\affiliation{School of Mathematics and Physics, China University of Geosciences, Wuhan 430074, China}

\author{Yang Liu}
\affiliation{Institute of Modern Physics, Chinese Academy of Sciences, Lanzhou 730000, China}
\affiliation{School of Nuclear Science and Technology, University of Chinese Academy of Sciences, Beijing 100049, China}

\author{Ye Tian}
\affiliation{Institute of Modern Physics, Chinese Academy of Sciences, Lanzhou 730000, China}
\affiliation{School of Nuclear Science and Technology, University of Chinese Academy of Sciences, Beijing 100049, China}

\author{Xionghong He}
\affiliation{Institute of Modern Physics, Chinese Academy of Sciences, Lanzhou 730000, China}
\affiliation{School of Nuclear Science and Technology, University of Chinese Academy of Sciences, Beijing 100049, China}

\author{Hao Qiu}
\affiliation{Institute of Modern Physics, Chinese Academy of Sciences, Lanzhou 730000, China}
\affiliation{School of Nuclear Science and Technology, University of Chinese Academy of Sciences, Beijing 100049, China}

\author{Xurong Chen}
\affiliation{Institute of Modern Physics, Chinese Academy of Sciences, Lanzhou 730000, China}
\affiliation{School of Nuclear Science and Technology, University of Chinese Academy of Sciences, Beijing 100049, China}


\date{\today}

\begin{abstract}
We present a simulation study of dark photon sensitivity in a suggested 
$\eta$ factory experiment at Huizhou. 
The vast number of $\eta$ mesons are produced by bombarding the high-intensity 
HIAF proton beam on a multi-layer target of light nucleus. 
The kinematic energy of the beam is at 1.8 GeV. 
For a prior experiment of one-month running, about $5.9 \times 10^{11}$ 
$\eta$ samples would be collected, 
providing substantial data for a sound statistical analysis. 
A compact spectrometer based on the full silicon-pixel tracker is conceptually 
designed for the detection of the final-state particles. 
The GiBUU event generator is utilized for the background 
estimation without the dark photon signal. 
A spectrometer simulation package ChnsRoot is constructed for evaluating 
the spectrometer performances in searching the dark photon in $\eta$ rare decay. 
The efficiency and resolutions of $\eta \rightarrow \gamma e^+e^-$ decay channel are studied in detail. 
The branching-ratio upper limit of dark photon in $\eta \rightarrow \gamma e^+e^-$ decay 
and the sensitivity to the model parameter are given from the simulations. 
\end{abstract}

\maketitle


\section{Introduction}
\label{sec:intro}

 The Standard Model (SM) of particle physics, despite being the most successful theory of fundamental particles and their interactions, is widely recognized as incomplete~\cite{Essig:2013lka, Bertone:2004pz}. Notably, the entire dark sector (DS) is not included in SM, 
and the broad range of SM parameters are not well understood~\cite{Bertone:2004pz}. 
It does not provide explanations of dark matter, dark energy observed 
in astrophysics and some anomalous signals in accelerator experiments,
such as the muon $(g-2)_{\mu}$ anomaly~\cite{ParticleDataGroup:2012pjm} 
and the angular correlation of electron and positron in $^8$Be decay~\cite{Krasznahorkay:2015iga, Krasznahorkay:2019lyl}. 

To explain such anomalies and the nature of dark matter, new theoretical constructs have been proposed, particularly involving light portal particles that mediate interactions between the SM and a hidden dark sector~\cite{Essig:2013lka, Pospelov:2007mp}. The dark photon is the earliest and most widely studied candidate for such a portal, proposed to interact with the SM via kinetic mixing~\cite{Holdom:1985ag, Fayet:2007ua} with the ordinary photon.

The decay process $\eta \rightarrow \gamma A (e^+e^-)$, where $A$ represents a dark photon, is of particular interest in particle physics, offering crucial insights~\cite{Beranek:2012ey, Raggi:2014zpa} into symmetries and the nature of dark photons in the mass range of the $\eta$ meson where experimental data is sparse. The mass of the $\eta$ meson and the kinematics of its decays make this channel well-suited for probing dark photon interactions with high sensitivity~\cite{Beranek:2013yqa}. Dark photons are believed to mediate interactions within the dark matter sector, much like how photons mediate electromagnetic interactions in the SM~\cite{Tulin:2017ara, Pospelov:2007mp}. Extensive theoretical and experimental work has focused on studying dark photons ($A$)~\cite{Essig:2013lka, Alexander:2016aln, Battaglieri:2017aum}.

Since dark photons are hypothesized to mediate interactions within the dark matter sector, understanding their production and decay mechanisms could shed light on the properties of dark matter itself. The search for dark photons is thus closely tied to broader efforts aimed at uncovering the nature of dark matter, one of the most pressing unsolved problems in modern physics.

While dark photons do not couple directly to SM particles, they can acquire a small coupling to the electromagnetic current through kinetic mixing between the SM hypercharge and the dark photon field strength tensors~\cite{Pospelov:2007mp, Arkani-Hamed:2008hhe, Bjorken:2009mm}. This coupling, suppressed by a factor denoted as $\epsilon$, opens a ``portal" that allows dark photons to be produced in laboratory experiments and subsequently decay into observable SM particles~\cite{LHCb:2019vmc, Graham:2021ggy}. If the kinetic mixing is induced via one- or two-loop processes involving heavy particles (potentially up to the Planck scale), the parameter $\epsilon^2$ is expected to range between $10^{-12}$ and $10^{-4}$~\cite{Alexander:2016aln}.

 In this study, a dark photon mass of 50 MeV has been used for the simulation of the channel $\eta \rightarrow \gamma A (e^+ e^-)$, which explores part of the higher mass region that remains largely unexplored. One of the key near-term goals of the dark-sector physics is to investigate this few-loop $\epsilon$ region.

Dark photons will decay into detectable~\cite{Tulin:2017ara} SM particles if decays into invisible dark-sector particles are kinematically prohibited. Constraints on such visible decays of dark photons have been set by earlier beam-dump experiments [e.g., E774~\cite{Bross:1989mp}, E141~\cite{Riordan:1987aw}, NuCal~\cite{Blumlein:2013cua}, Charm~\cite{Gninenko:2012eq}], collider/fixed target experiments [A1~\cite{Merkel:2014avp}, LHCb~\cite{LHCb:2019vmc}, NA48/2~\cite{NA482:2015wmo} and rare meson decay~\cite{KLOE-2:2011hhj, KLOE-2:2012lii, WASA-at-COSY:2013zom, HADES:2013nab, PHENIX:2014duq, KLOE-2:2016ydq} experiments]. 
These experiments have excluded the few-loop region for dark photon masses $m(A) \leq 10$ MeV, but much of the parameter space at higher masses remains unexplored. This gap exists due to limitations in experimental sensitivity, particularly at the higher energy ranges. The current study aims to address this by simulating dark photon production, assuming a fixed dark photon mass of $50$ MeV, which remains largely unprobed. By exploring this higher mass region, we seek to provide critical insights for future experimental searches.
 
A proposed super eta factory in Huizhou, leveraging high-intensity proton beams from HIAF or CiADS, offers a promising platform to study dark photon production and decay~\cite{Chen:2024wad}. With its ability to produce high-intensity beams and access unexplored mass regions, this facility could play a pivotal role in advancing our understanding of dark sector physics. In this work, we present preliminary simulation results for dark photon searches at the proposed eta factory, laying the groundwork for future experimental investigations.

\section{Theoretical model of dark photon}
\label{sec:theory}

\subsection{Introduction to Vector Portal Model}
The vector portal model is a theoretical framework that proposes a new type of interaction between standard model particle and a new hidden sector often referred to as a ``dark sector''. This interaction is mediated by a new gauge boson, often called a dark photon ($A$), which is analogous to the photon in the Standard Model. The Vector Portal offers different types of model categories. The model currently under study is minimal dark photon model, which serves as the simplest realization of the vector portal. This dark photon acquires mass through a hidden mechanism, such as the Stueckelberg mechanism~\cite{Ruegg:2003ps} or a dark Higgs field. The corresponding Lagrangian for the minimal dark sector~\cite{Lanfranchi:2020crw} is given by: 
\begin{align}
\mathcal{L} = & -\frac{1}{4} F_{\mu\nu} F^{\mu\nu} - \frac{1}{4} F'_{\mu\nu} F'^{\mu\nu} 
- \frac{\epsilon}{2} F_{\mu\nu} F'^{\mu\nu} \nonumber \\
& + \frac{1}{2} m_{A'}^2 A'_{\mu} A'^{\mu} + A'_{\mu} J_{\text{dark}}^{\mu} 
+ \mathcal{L}_{\text{mass}} 
 \label{eq:lag}
\end{align}

In Eq.~\eqref{eq:lag}, $F_{\mu\nu}$ is the field strength tensor for the Standard Model photon, while $F'_{\mu\nu}$ represents the field strength tensor of the dark photon field. The term $\epsilon$ is the kinetic mixing parameter, which quantifies the coupling between the standard model photon and the dark photon. The mass of the dark photon is denoted by $m_{A'}$, and it arises from the introduction of a dark Higgs or Stueckelberg mechanism, both of which break the $U(1)_A$ symmetry. The dark photon field itself is represented by $A'_{\mu}$, while the interaction between the dark photon and the dark sector is given by the current $J_{\text{dark}}^{\mu}$. This current describes interactions with new fermions or scalars in the dark sector.

Moreover, $\mathcal{L}_{\text{mass}}$ includes terms related to mass generation for the dark photon and possible contributions from a dark Higgs mechanism, as well as any spontaneous symmetry-breaking effects that may lead to mass splitting of dark sector fields.

Together, these terms in the Lagrangian describe the dynamics and interactions of both the electromagnetic and dark sectors, including the possibility of new physics interactions mediated by the dark photon.

\subsection{Minimal dark photon model}
One of the most important models in the vector portal models is \emph{Minimal dark photon model}. The SM in this scenario is being extended by introducing the concept of ($A$). It interacts with regular matter through a connection called kinetic mixing $\epsilon^{2}$~\cite{Holdom:1985ag}. Simulation at HIAF is done to observe new particles in the decays $\eta \rightarrow \gamma A (e^+ e^-)$.

\begin{figure}[hbt!]
\centerline{\includegraphics[width=0.9\linewidth]{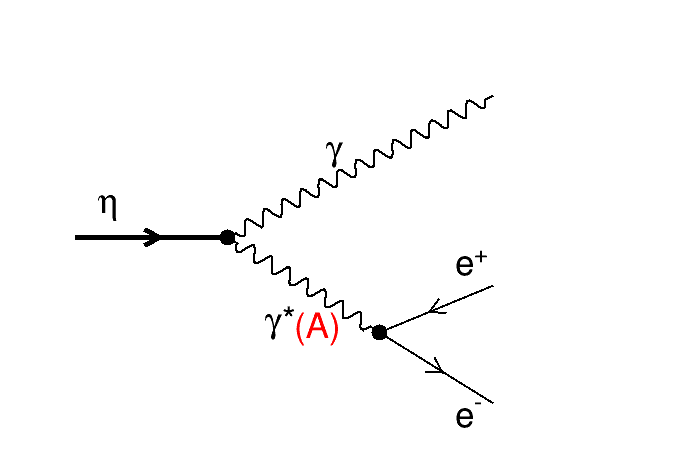}}
\caption{Schematic diagram of $\eta$ decay into $\gamma A$ via a virtual photon $\gamma^*$, where $A$ denotes the dark photon. The final state includes an electron and a positron. $\eta \rightarrow \gamma e^{+}e^{-}$ is the main background for dark photon search.}
\label{fig:fynman}
\end{figure}

$\eta \rightarrow \gamma e^+e^-$ has comparatively high branching ratio i.e. $0.69\%$~\cite{ParticleDataGroup:2016lqr}. The minimal models assume that the dark photon decays entirely in a visible manner~\cite{BaBar:2014zli, Riordan:1987aw, Merkel:2014avp, NA482:2015wmo} or invisibly into dark matter particles~\cite{Izaguirre:2015yja, Fayet:2004bw}. The simulation results indicate that HIAF will be capable of detecting the final state particles with sample size larger than $10^8$. Globally, many other experiments are performing Dark photon searches like KLOE-2, Thomas Jefferson National Accelerator Facility (JLAB)~\cite{HPS:2018xkw}, Beijing Electron Spectrometer III (BESIII)~\cite{BESIII:2024pxo}, Positron Annihilation into Dark Mediator Experiment (PADME)~\cite{Raggi:2015gza}, and the proposed Rare Eta decays To Observe Physics Beyond the Standard Model (REDTOP)~\cite{REDTOP:2022slw}. The preliminary sensitivity studies have been performed to understand how effectively the HIAF can detect signals predicted by the \emph{minimal dark photon model}.

\section{Compact pixelated silicon spectrometer}
\label{sec:simulation}

\begin{figure}[h]
\centerline{\includegraphics[width=0.9\linewidth]{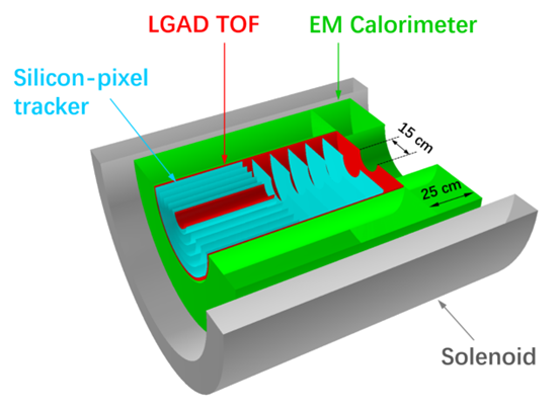}}
\caption{The conceptual design of compact spectrometer of $\eta$ factory. The spectrometer mainly relies on the silicon detector technology, with a monolithic silicon pixel tracker and the Fast LGAD TOF detector of low material budget. The silicon tracker is wrapped with a fast lead-glass calorimeter for high-energy photons.}
\label{fig:PSTD}
\end{figure}

The decay process $\eta \rightarrow \gamma e^+ e^-$ which is measured using a pure silicon track detector (PSTD) a conceptual designed spectrometer~\cite{An:2025lws} shown in Fig.~\ref{fig:PSTD} for the proposed $\eta$ factory at HIAF. The PSTD has several advanced features. It is an all silicon tracker, optimized to handle ultra high event rates exceeding 100 MHz, which allows for fast collection of statistics, exceeding current global rate for $\eta$ physics. Its compact design integrates a tracker and Low Gain Avalanche Diode (LGAD) based Time-Of-Flight (TOF) system within approximately a 30 cm radius, which significantly reduces costs for the calorimeter, magnet, and muon detector.

It provides high position resolution, which is crucial for meeting momentum resolution requirements even with shorter track lengths. The spectrometer is designed to minimize background for reconstructed particles with secondary vertices, thus enhancing the detection of hyper-nuclei. With a focus on efficient data handling, the spectrometer is capable of distinguishing hits from different events at a 100 MHz event rate, demonstrating its strength in high-density environments. Furthermore, it features a pixel tracker with a radius ranging from 7.5 to 27.5 cm, 5 barrel layers, and 5 disks, enhancing its tracking capabilities.

Both lead glass and a dual-readout design combining lead glass with plastic scintillator are being considered as candidate technologies for electromagnetic calorimeter R\&D. When a particle passes through lead glass, it produces only Cherenkov light without generating scintillation light, which effectively suppresses background from neutrons and pions. However, due to the low Cherenkov light yield of lead glass, the achievable energy resolution is limited to around 6$\%/\sqrt{\text{GeV}}$. To improve the energy resolution, a dual-readout approach inspired by the ADRIANO2\cite{instruments6040049} design is being considered, in which lead glass and plastic scintillator are arranged in a longitudinally layered structure. This configuration is expected to achieve an energy resolution of approximately 3$\%/\sqrt{\text{GeV}}$.

For the future experiments at HIAF, the statistics of $\eta$-meson samples play a critical role. The statistics depend on the energy configuration, cross-section, and duration of the future experiment. The proton beam energy is set at 1.8 GeV, slightly below the $\rho$ production threshold, to minimize background. At this energy, the $\eta$ production probability in elastic scattering is approximately 0.76$\%$, as determined by Giessen Boltzmann-Uehling-Uhlenbec (GiBUU) simulations of $p$-$^{7}\text{Li}$ collisions. Based on previous measurements, the $\eta$ production cross section in p-p collisions at 1.8~GeV is approximately 0.1~mb. Consequently, the cross section in $p$-$A$ collisions is approximately $0.1 \times A$ mb. The experimental setup utilizes a multi-layer target of light nuclei (lithium) to enhance $\eta$ meson production and minimize background and particle multiplicity as the multi-layer target of thin foils (Lithium or Beryllium) will be used in the future experiment. The luminosity of the fixed target experiment is expected to reach $10^{35} \, \text{cm}^{-2} \text{s}^{-1}$. Considering the event rate capacity of the spectrometer, we adopt a conservative estimate for the inelastic scattering event rate at 100 MHz. To evaluate the potential impact of the future experiment, we assume it will operate for one month with a duty factor of 30$\%$. Under these assumptions, the total number of $\eta$-mesons produced in a prospective experiment is estimated to be $5.9 \times 10^{11}$.

Due to computational constraints, this study simulated approximately $1.3 \times 10^{7}$ inelastic $~pA$ collision events. To estimate the sensitivity of the actual experiment, both the background distributions and the number of produced 
$\eta$-mesons are scaled up using a factor of approximately $10^{5}$. The future $\eta$-factory experiment is expected to achieve an impressive total event count.

\section{Simulation Framework}
\label{sec:simulation}
To explore the potential of detecting dark photon particles at HIAF, we executed a simulation using the event generator GiBUU, where the proton beam interacts with a $^7Li$ target, facilating the production of $\eta$ particles.

The simulation studies for the decay process $\eta \rightarrow \gamma A (e^+ e^-)$ were conducted using a conceptual designed spectrometer. The GiBUU event generator was used to simulate the $\eta$ meson production and decay process, while the ChnsRoot package was employed for detector response and event reconstruction. These simulations aim to validate the effectiveness of the spectrometer for future operations at HIAF. The GiBUU event generator is based on the Boltzmann and Uehling-Uhlenbeck equations~\cite{Buss:2011mx,Gaitanos:2007mm,Weil:2012ji}, which model particle interactions and transport mechanisms in nuclear environments. Its applications cover a broad energy range (100 MeV - 100 GeV) and include various nuclear physics phenomena such as heavy-ion collisions, nuclear reactions, and $\eta$-meson production. Using the GiBUU event generator for $\eta$-meson production , we programmed specific decay chains of the $\eta$-meson to create a more realistic simulation and to analyze the efficiency and resolution of the decay channel.

ChnsRoot package was developed for the spectrometer simulation which is based on FairRoot framework. The FairRoot framework provides essential tools for detector simulation and offline analysis, allowing users to design experimental setups efficiently and with ease. In ChnsRoot, we implement a fast simulation based on the results from Geant4 simulations, which provide insights into the energy resolution and efficiency of the detectors. Using the ChnsRoot package, we are able to study the detector acceptances, efficiencies, and resolutions both efficiently and reliably.

\begin{figure*}[hbt!]
\centerline{\includegraphics[width=0.8\linewidth]{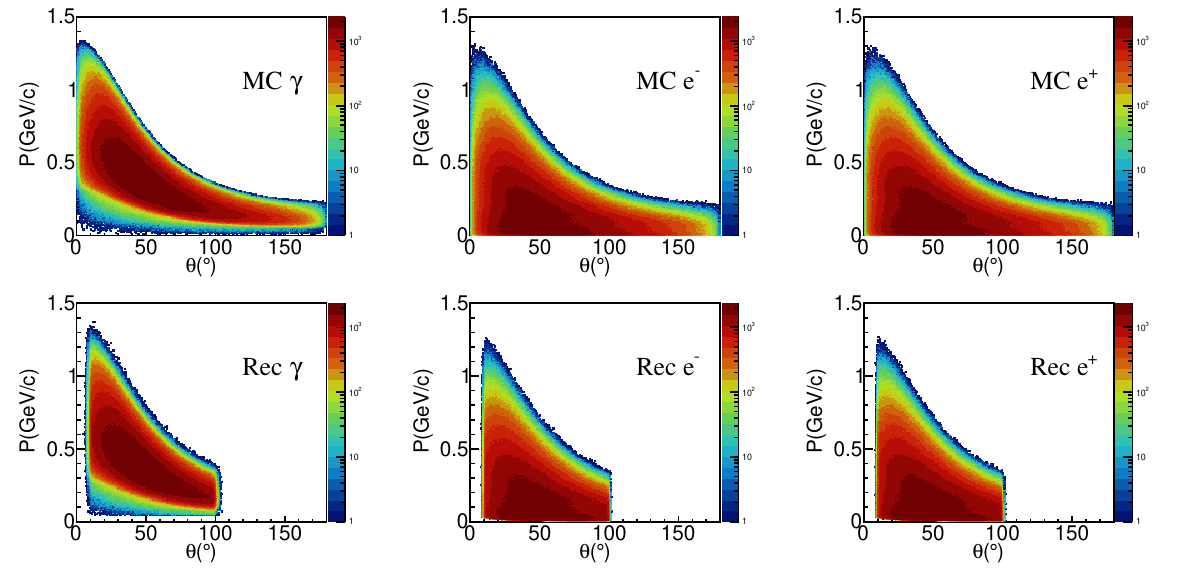}}
\caption{The distribution of momentum to angle for the final state particles . The upper half of a figure represents reconstruction obtained from a simulated detector, while the lower half represents generation through Monte Carlo simulation.}
\label{fig:Kinetics}
\end{figure*}

The angular acceptance of the spectrometer is designed to detect particles within $10^{\circ} - 100^{\circ}$. Photons with energies as low as 50 MeV can be reliably detected, while the silicon pixel tracker effectively captures charged particle tracks. The reconstructed kinematics of the $\eta$-meson decay products (photon, electron, and positron) are shown in Fig.~\ref{fig:Kinetics}, illustrating their momentum distribution. The inner radius of the silicon pixel tracker imposes a threshold for detecting the minimum momentum of charged particles. The EMC hit threshold for neutral particles is set to the energy equivalent of a 50 MeV photon. Simulations show that this EMC threshold effectively removes low-energy neutron background while retaining as many photons as possible.

\section{Results and Discussions}

Based on the Monte Carlo simulations, we first estimate the detection efficiencies for the $\eta \rightarrow \gamma e^+e^-$. Next, we present the mass resolution of the $\eta$ meson and the $e^{+} e^{-}$ pair. Then, we display the projected background distributions after applying the event selection criteria. Subsequently, we calculate the upper limits of the branching ratios for $\eta \rightarrow \gamma e^+e^-$. Finally, the sensitivity to the model parameters is derived from the simulation data.
\subsection{Efficiency}

\begin{figure}[hbt!]
\centerline{\includegraphics[width=0.9\linewidth]{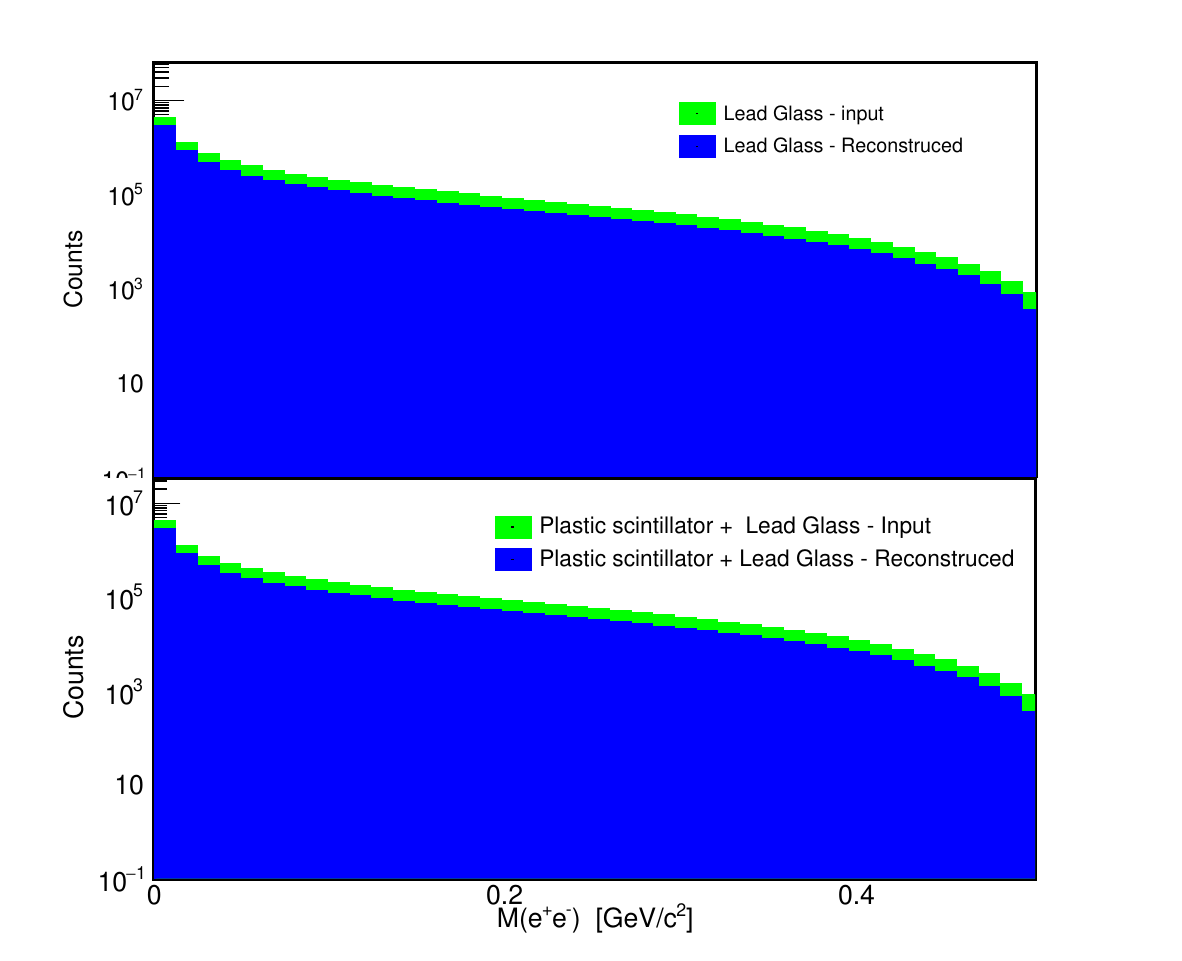}}
\caption{The event distributions as a function of dilepton mass spectrum for the channel $\eta \rightarrow e^{+}e^{-}\gamma$ are presented. The green histogram represents the input Monte Carlo events generated by the event generator, while the blue histogram depicts the reconstructed events from the detector simulation.}
\label{fig:finalratio}
\end{figure}

\begin{figure}[hbt!]
\centerline{\includegraphics[width=0.9\linewidth]{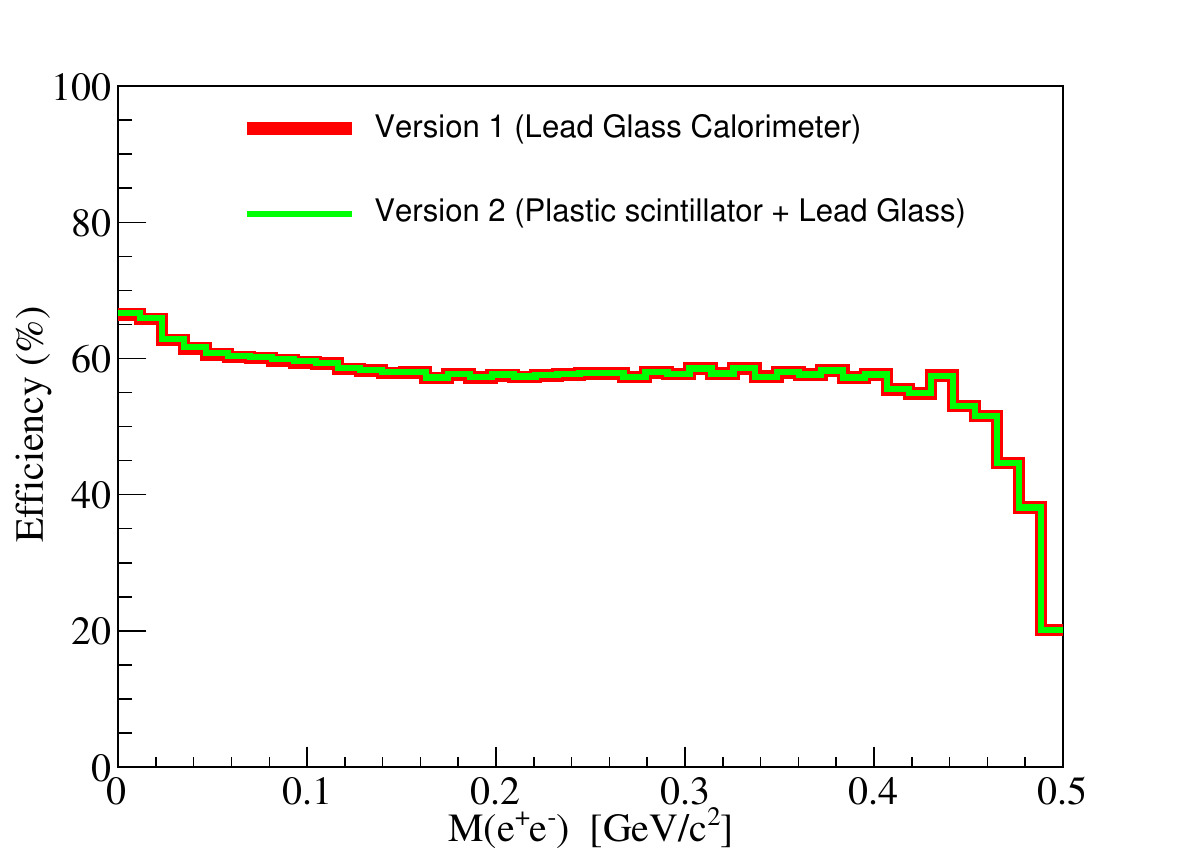}}
\caption{A graph displaying the efficiency of the channel $\eta \rightarrow e^{+}e^{-}\gamma$ as a function of mass of $e^{+}e^{-}$, illustrating the efficiency stability across different mass ranges.}
\label{fig:Efficiency}
\end{figure}

The detection efficiency for the targeted $\eta$ decay channel is a crucial factor in optimizing the spectrometer design. Accurate measurement of efficiency helps ensure that the spectrometer is capable of identifying the decay products with high precision. Fig.~\ref{fig:finalratio} compares the invariant mass spectra  of $e^{+}e^{-}$ pairs of events generated by the Monte Carlo (MC) event generator GIBUU and those MC reconstructed. The x-axis represents the invariant mass in $GeV/c^{2}$,  while the y-axis shows the event counts on a logarithmic scale. Both distributions exhibit a higher count of events at lower invariant masses, decreasing gradually at higher masses.
Fig.~\ref{fig:Efficiency} shows the efficiency value which is computed by taking the ratio of the content of two histograms, one representing data from event generator and other from reconstructed data. The efficiency is calculated for each bin. The overall efficiency for the channel is estimated to be $\sim$ $60\%$. This efficiency is quite satisfactory, as it closely aligns with the pure geometrical acceptance.

\subsection{Resolutions}

\begin{figure}[hbt!]
\centerline{\includegraphics[width=0.9\linewidth]{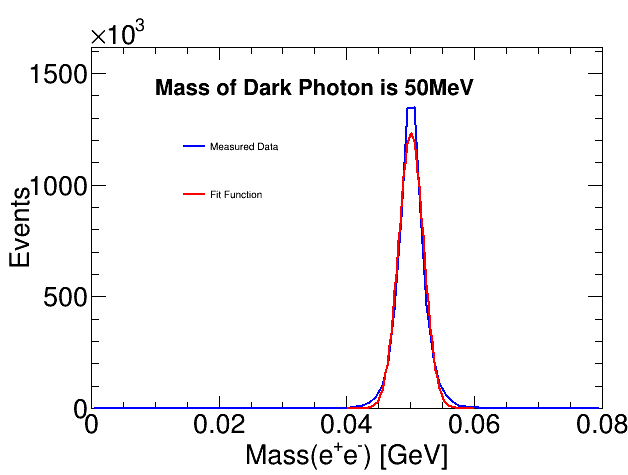}}
\caption{The fit to the $M(e^{+}e^{-})$ distribution for $\eta \rightarrow e^{+} e^{-} \gamma$. The blue line represents data and the red line is the total fit at the mass of dark photon 50 MeV assumed in the simulation.}
\label{fig:MC_gamma}
\end{figure}

Fig.~\ref{fig:MC_gamma} shows the resolution of the $e^{+}e^{-}$ pair resulting from the decay of a dark photon. The peak at 50 MeV, corresponding to the dark photon mass, merges with GIBUU simulated background. Gaussian fit was performed, and signal width was defined as $6\sigma$. By fitting the dark photon signal, we determined the signal width $\sigma_{e^{+}e^{-}}$  is $1.94$ MeV and the number of bins required for the analysis. The high spatial resolution of the silicon pixel detector results in a mass resolution for the dark vector particle of less than 2 MeV. The small mass resolution is essential for the sensitivity to new particles, as it reduces the number of background events beneath a narrower peak.  The close agreement between the measured data and the fit function indicates the accuracy of our reconstruction process. This mass reconstruction is crucial for reducing uncertainties in the measurement of the mixing parameter
$\epsilon^{2}$, there by improving our sensitivity study.

\begin{figure}[hbt!]
\centerline{\includegraphics[width=0.9\linewidth]{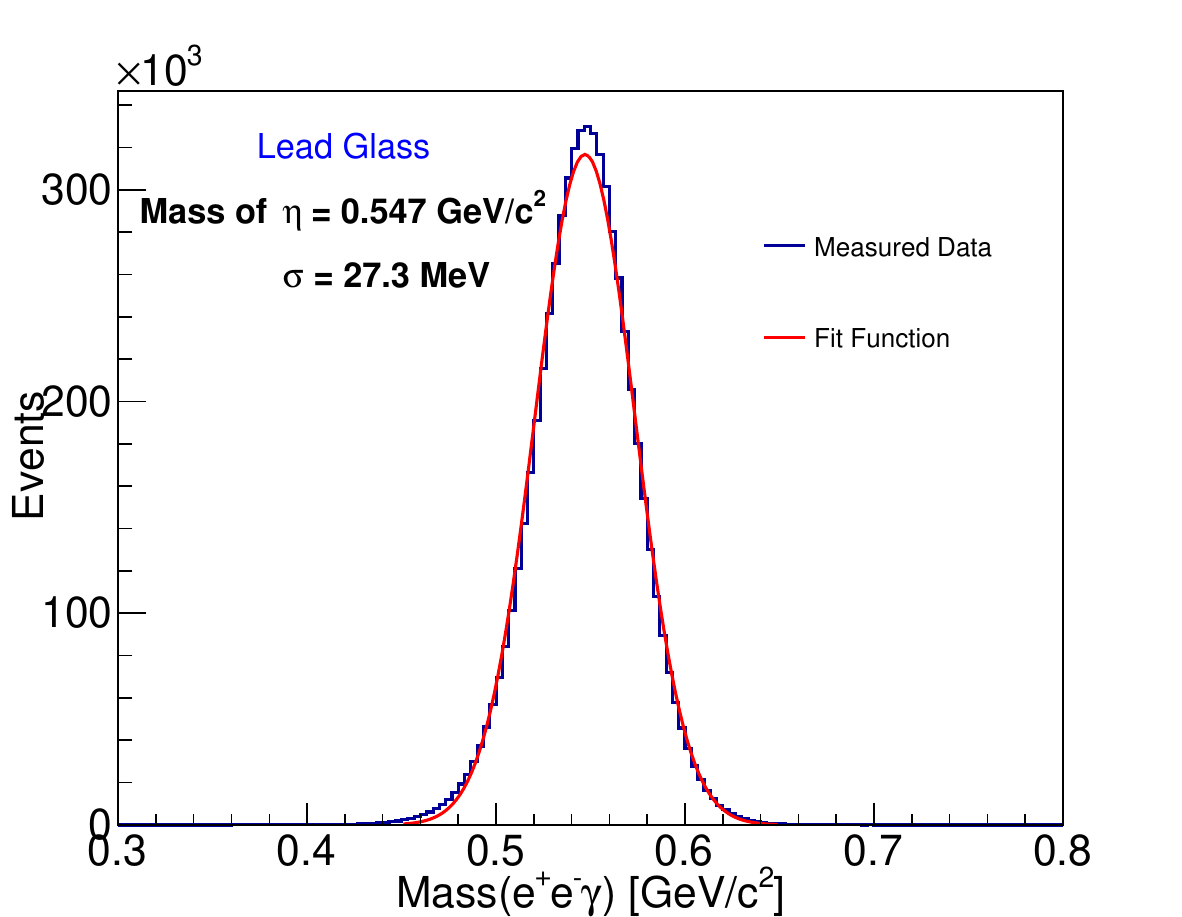}}
\caption{The fit to the $M(e^{+}e^{-}\gamma)$ distribution for $\eta \rightarrow e^{+} e^{-} \gamma$ with version 1 (lead glass). The blue line represents data and the red line is the total fit at the mass of $\eta$ 0.547 GeV.}
\label{fig:MC_eta_v1}
\end{figure}

\begin{figure}[hbt!]
\centerline{\includegraphics[width=0.9\linewidth]{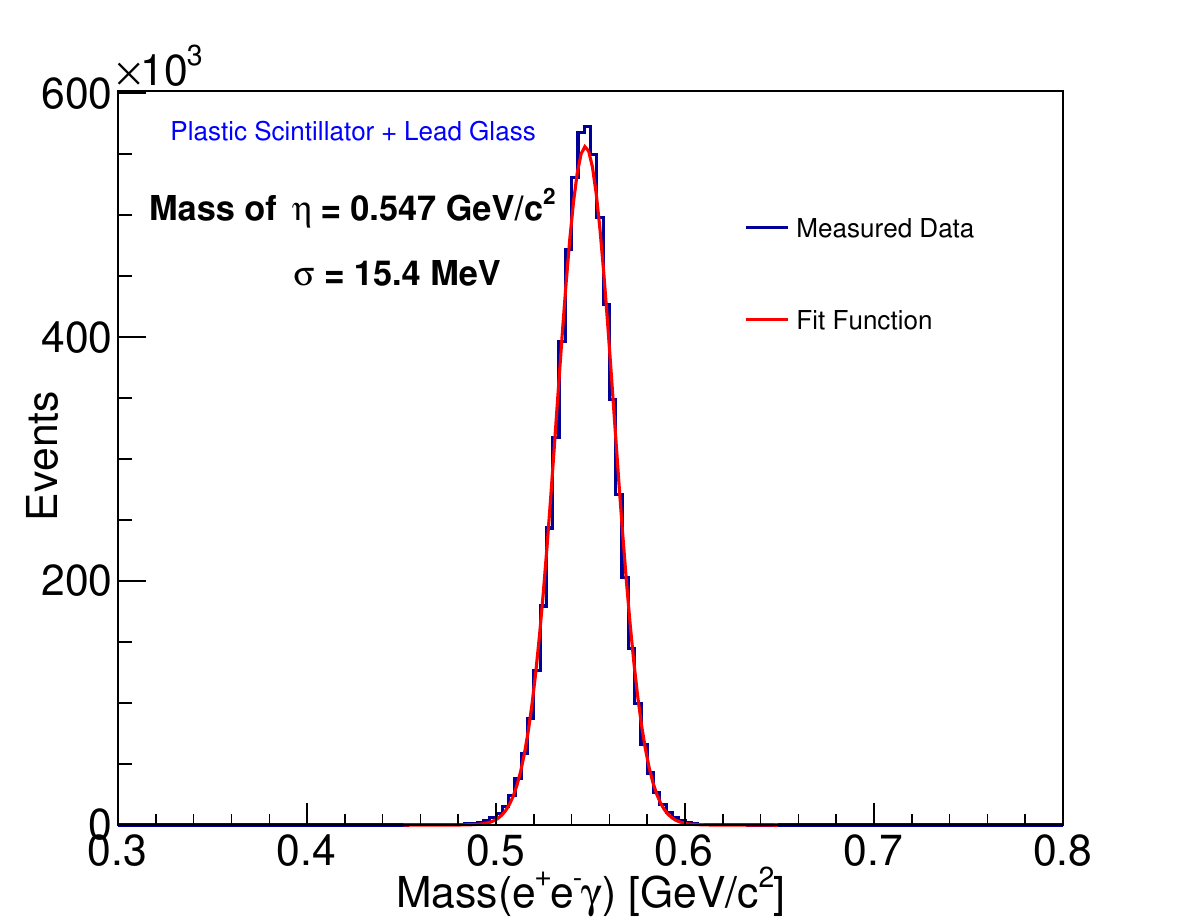}}
\caption{The fit to the $M(e^{+}e^{-}\gamma)$ distribution for $\eta \rightarrow e^{+} e^{-} \gamma$with version 2 (lead glass + plastic scintillator). The blue line represents data and the red line is the total fit at the mass of $\eta$ 0.547 GeV.}
\label{fig:MC_eta_v2}
\end{figure}

 For the photon energy cut, we set a threshold where the energy of the candidate photon must be greater than $0.05$ GeV for it to be considered a valid hit to the electromagnetic calorimeter (EMCAL). 

In Fig.~\ref{fig:MC_eta_v1} and Fig.~\ref{fig:MC_eta_v2}, we compare the resolution of ChnsRoot version 1 (v1) and ChnsRoot version 2 (v2) for both the Lead Glass Calorimeter and a setup with alternating layers of lead glass and plastic scintillator. The sigma values of the resolution curves show a significant improvement in v2. For v1, the sigma is 27.3 MeV, whereas v2 achieves a much smaller sigma value of 15.4 MeV. This represents a $43\%$ reduction in sigma, indicating that v2 offers a considerably sharper resolution, enhancing the ability to detect and differentiate signals with higher precision. The improved resolution in v2 suggests better overall performance making it more reliable for experiments requiring high precision and minimal noise. 
 The peak around 0.55 GeV/$c^{2}$ validates the presence of the expected decay products. This reconstruction enables precise determination of the mixing parameter $\epsilon^{2}$, thereby enhancing the overall sensitivity and reliability of our results.
 
 In addition to the kinematic distribution of the final state particles, it is necessary to apply selection criteria to these particles to filter out excess background noise and accurately reconstruct the $\eta$ meson. While selecting the decay channel $\eta \rightarrow e^{+}e^{-}\gamma$ to reconstruct the $\eta$ meson, we plot the invariant mass distribution of $e^{+} e^{-} \gamma$. However, this method is susceptible to significant background interference. In order to suppress the background, specific cuts on $\eta$ mass and photon energy cuts were applied. For lead glass calorimeter, the mass cut is $0.49 < M_{\gamma e^{+} e^{-}} < 0.61$. For lead glass + plastic scintillator setup, a slightly tighter mass cut of $0.5 < M_{\gamma e^{+} e^{-}} < 0.6$ is used.

In our study, the detector cannot identify the difference between photons and neutrons, which can lead to errors. To fix this, we adjust the properties of neutrons to make them appear as photons to the detector by changing their mass to zero, similar to photons. This method enables us to assess how this limitation impacts our findings more accurately.

\subsection{Background distributions}
\begin{figure}[hbt!]
\centerline{\includegraphics[width=0.9\linewidth]{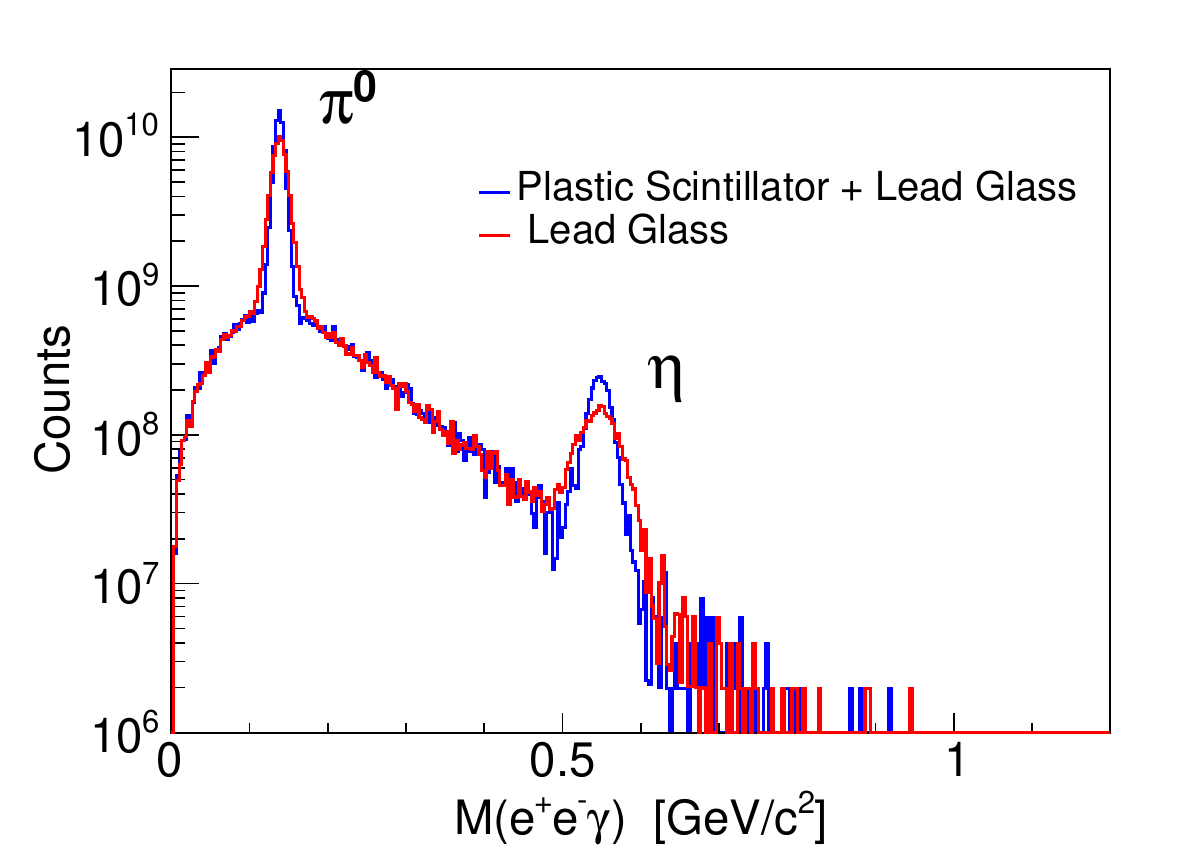}}
\caption{The projected invariant mass distribution of $\gamma e^{+} e^{-}$ with the background from GiBUU based on the proposed one-month yield of $\eta$.}
\label{fig:final_eta}
\end{figure}

\begin{figure}[hbt!]
\centerline{\includegraphics[width=0.9\linewidth]{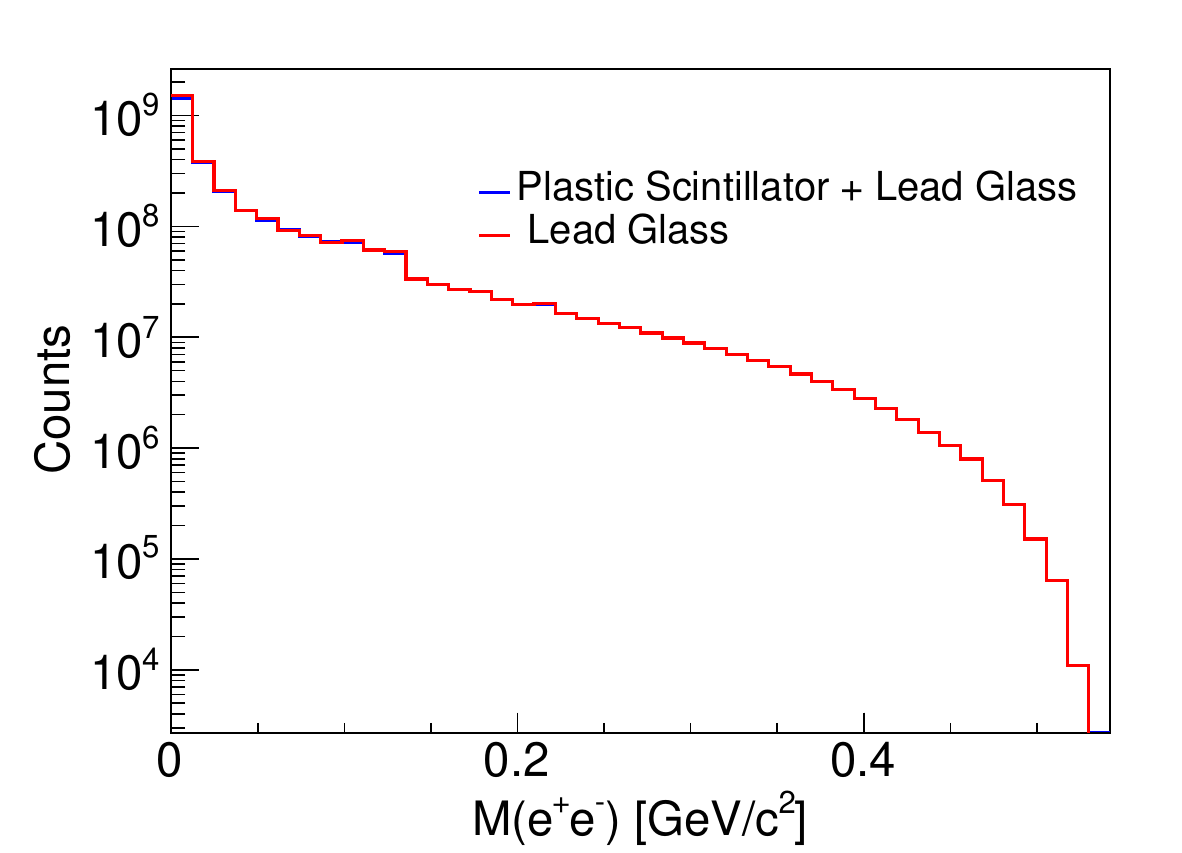}}
\caption{The projected invariant mass distribution of $e^{+} e^{-}$ in the channel $\eta \rightarrow e^{+}e^{-} \gamma$ is based on the proposed one-month run time of the experiment.}
\label{fig:bothdil}
\end{figure}

The decay channel $\eta \rightarrow e^{+}e^{-}\gamma$ is focused for the dark photon search. The method involves identifying a bump in the invariant mass distribution of $e^{+}e^{-}$. Before generating the desired invariant mass distributions, the channel of interest was selected. The reconstructed $\eta$ masses are required to fall within $\pm 3\sigma$. Figures~\ref{fig:final_eta} and Figure~\ref{fig:bothdil} depict the simulated invariant mass distributions for $e^{+}e^{-}\gamma$ and $e^{+}e^{-}$, respectively. For the histogram $e^{+}e^{-}$, the number of bins was determined by the relationship $2m_e < m_A < m_\eta - m_A$, where $m_\eta = 0.547$ GeV, $m_A = 0.05$ GeV. Subtracting $m_A$ from $m_\eta$ gives a mass range of 0.497 GeV. Using a bin width of approximately six times the dark scalar particle resolution ($6 \times 0.001943$ GeV), the number of bins is calculated as
\[
\frac{0.497}{0.011658} \approx 42.63,
\]
which is rounded to 43 bins. This selection ensures adequate resolution for analyzing the signal within the chosen range. For conservative estimation in the detector simulation, neutrons with energy above the EMC hit threshold are treated as misidentified photons. 

The Figure~\ref{fig:bothdil} is the comparison of $e^{+}e^{-}$ invariant mass distributions generated by both versions of ChnsRoot. The invariant mass of $e^{+}e^{-}$ was restricted to a mass window of $0.49 < M_{\gamma e^{+} e^{-}} < 0.61$ for v1 and $0.5 < M_{\gamma e^{+} e^{-}} < 0.6$ for v2 according to the $\gamma e^{+} e^{-}$. Since the dark vector particle is not included in the GiBUU event generator. As a result, the obtained invariant mass distribution represents only the background, excluding the potential signal of dark photon. Lower background levels generally enhance the sensitivity of the experiment.

Figure~\ref{fig:final_eta} shows the combination of background events simulated by the GiBUU event generator with the signal from $\eta$ meson decays. Background events were subject to both photon energy and mass cuts. These mass and energy cuts helps to reduce the inclusion of background events and improves the purity of the signal, enhancing the sensitivity of the experiment to potential new physics.

\begin{figure}[hbt!]
\centerline{\includegraphics[width=0.9\linewidth]{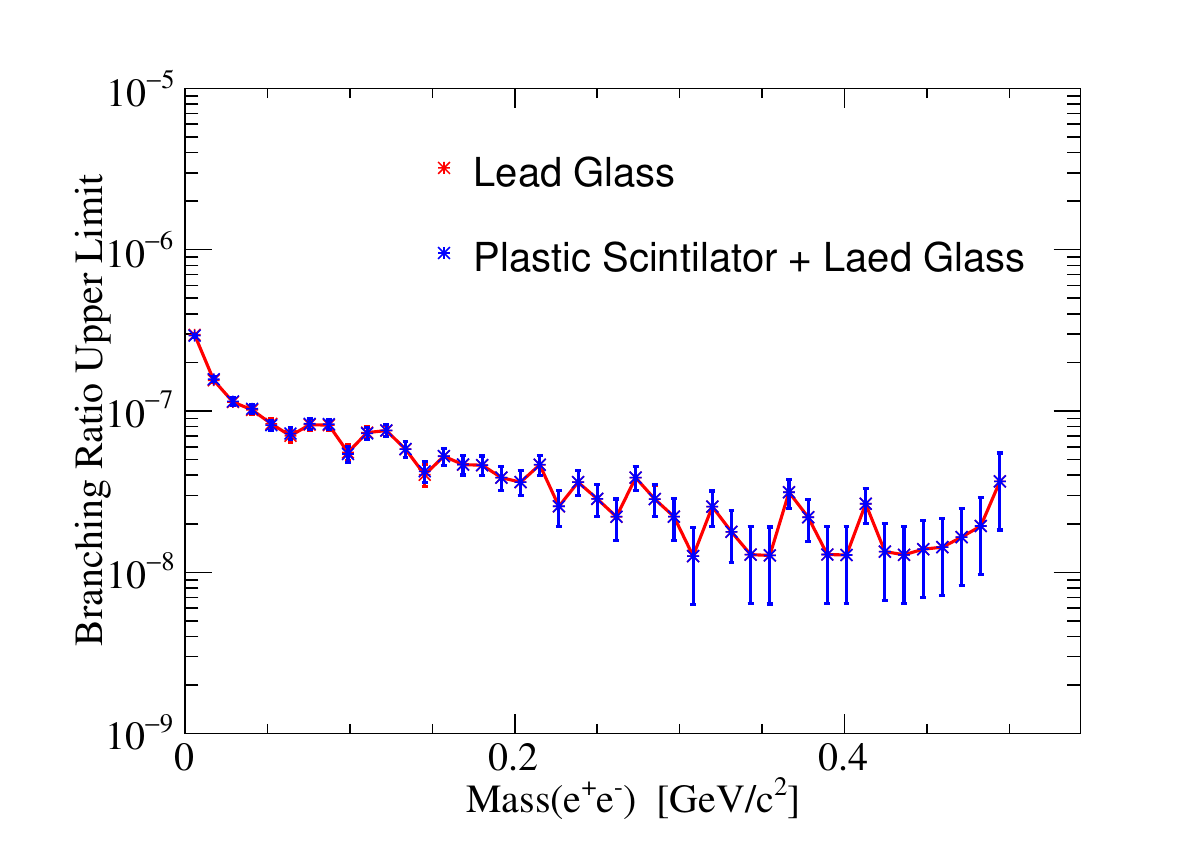}}
\caption{The projected branching ratio upper limit in the decay channel $\eta \to \gamma A \to e^+ e^- \gamma$ based on the proposed one-month run time of the experiment.}
\label{fig:BR}
\end{figure}

\subsection{Sensitivity Studies for Probing Physics Beyond
the Standard Model}

\subsubsection{Branching ratio upper limit}

In the search of dark photons, sensitivity studies are essential to establish limits on their possible existence and interaction strengths ~\cite{BaBar:2014zli}. By analyzing the invariant mass distribution of $e^{+}e^{-}$, we set an upper bound on the branching ratio for the dark photon decay channel, providing critical insights into the experimental sensitivity. Using the background distributions after event selection, branching ratio upper limit can be calculated based on the total number of $\eta$-mesons produced, the detection efficiency, and the background event count in each mass bin. The sensitivity to the branching ratio upper limit of the decay $\eta \to \gamma A \to e^+ e^- \gamma$, assuming a specific dark photon mass, has been evaluated by analyzing the invariant mass distribution of the $e^{+}e^{-}$ pair. The achieved sensitivity level is on the order of $10^{-7}$. The sensitivity `S' of branching ratio (Br) is the Br upper limit given by:
\begin{equation}
	\begin{split}
	    S(\text{Br}(\bar{g})) = \frac{3\times \sqrt{N_{\text{bkg}}^{i}}}{N_{\eta}\times\epsilon_{reco}}\;.
	\end{split}
	\label{eq:BR}
\end{equation}
Here, $\bar{g}$ represents the channel of interest, $N_{\text{bkg}}^{i}$  represents the number of background events and $N_{\eta}$ signifies the total number of $\eta$ produced. Additionally, $\epsilon_{reco}$ represents the reconstruction efficiency for the channel. The resulting branching ratio upper limit is shown in Fig.~\ref{fig:BR} for both versions of ChnsRoot, as a function of the invariant mass of the dark photon. Comparing the two versions analyzed, v2 offers a marginal improvement in sensitivity over v1. The error bars are statistical, with larger fluctuations at lower masses and improved precision at higher masses. The branching ratio upper limit decreases as the invariant mass increases from 0 to approximately 0.1 GeV$/c^{2}$, indicating enhanced sensitivity in this mass range. Beyond 0.1 GeV$/c^{2}$ the upper limit stabilizes, demonstrating consistent sensitivity across higher masses.

\subsubsection{Sensitivity to mixing parameter ($\epsilon^{2}$)}
In the context of rare decay channel $\eta \rightarrow e^{+}e^{-} \gamma$, sensitivity studies are performed to constrain the mixing parameter $\epsilon^{2}$ that governs the interaction strength between dark photon and the SM particles.
The sensitivity to the mixing parameter $\epsilon^{2}$~\cite{WASA-at-COSY:2013zom} is computed based on the upper limit of the branching ratio for the dark photon decay channel. The relationship is given by: 

\begin{equation}
	\begin{split}
		S(\epsilon^2) = \frac{S(Br(\bar{g}))}{2 |F(m_{A}^2)|^2 (1 - \frac{m_{A}^2}{m_{\eta}^2})^3}\;.
	\end{split}
	\label{eq:sensitivity}
\end{equation}

\begin{figure}[hbt!]
\centerline{\includegraphics[width=0.9\linewidth]{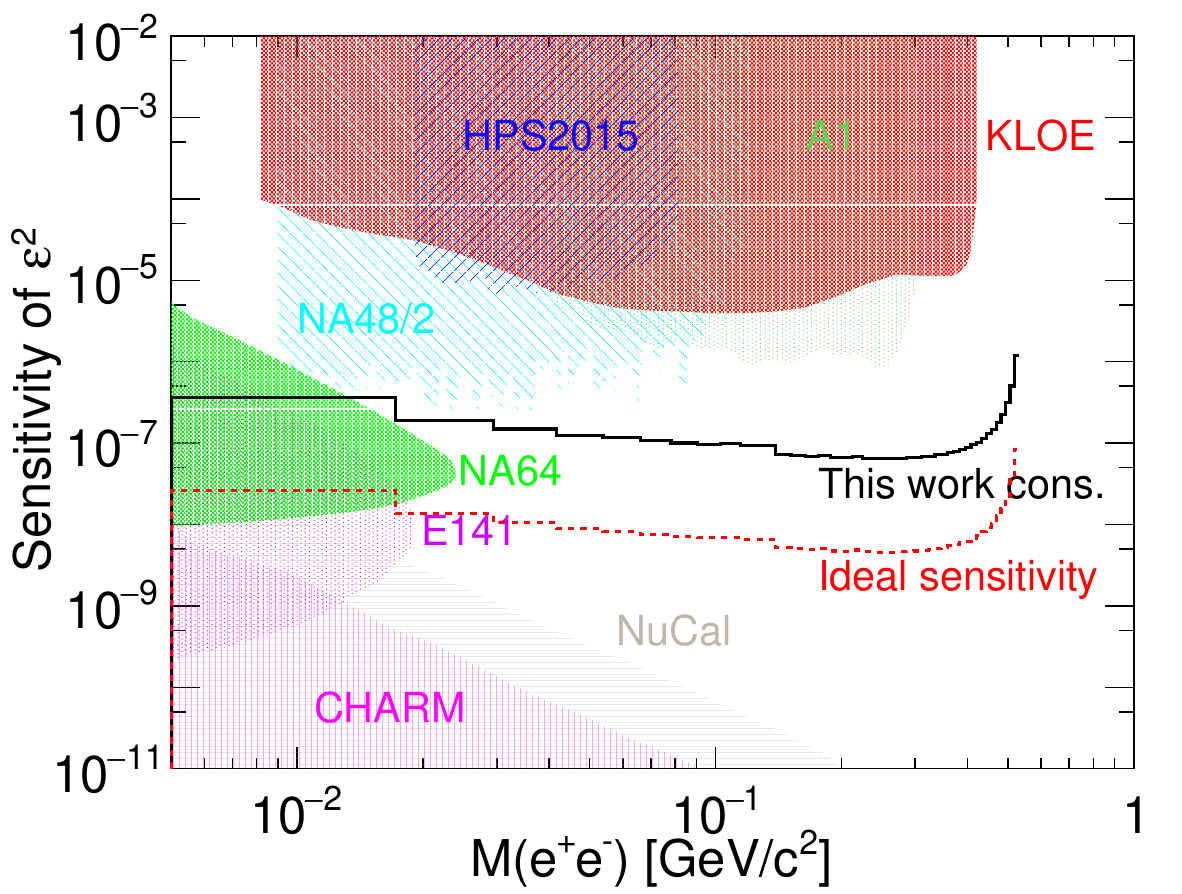}}
\caption{Sensitivity to $\epsilon^{2}$ mixing parameter as a function of $m_A$ for its decay into visible final state. The plot includes existing limits from previous searches and the black solid curve represents the sensitivity based on one month of data, while the red dotted curve shows the ideal projected sensitivity assuming one year of data taking at 500 MHz with 100\% duty factor at HIAF.}
\label{fig:Epsilon2}
\end{figure}

Where, $m_{A}$ is the mass of the dark photon, $F(m_{A})$ is the transition form factor (TFF) of the $\eta$ meson, and $m_{\eta}$ is the mass of the $\eta$ meson. The TFF of $\eta$ meson meson describes its internal structure and its off-shell coupling to photons (or dark photons). The observed rate for $\eta \rightarrow \gamma A$ also depends on the dark photon's kinetic mixing parameter $\epsilon$, which governs its coupling to the electromagnetic current~\cite{A2:2013wad}. By applying this formula, we can calculate the sensitivity to $\epsilon^{2}$ across different mass ranges of the dark photon, allowing us to determine the smallest values of $\epsilon^{2}$ that can be probed in the experiment. 

The sensitivity to the mixing parameter $\epsilon^{2}$ reflects the precision at which $\epsilon^{2}$ can be probed to test the vector portal model with meaningful significance. This sensitivity is intrinsically linked to the measured upper limit of the branching ratio for the channel under study, via Eq. \eqref{eq:BR}. By determining $\epsilon^{2}$ sensitivity, we establish how effectively the experiment can constrain the strength of interaction between dark photon and standard model particles.

Fig.~\ref{fig:Epsilon2} indicates the projected sensitivity to $\epsilon^{2}$ as a function of mass of dark photon, alongside comparisons to existing constraints from other experiments. For our proposed setup, the sensitivity reaches $\epsilon^{2}$ $\sim$ $10^{-7}$ at $99\%$ confidence level (CL) with one month of data. An extended run time would further improve sensitivity and tighten constraints on dark photon parameters. Fig.~\ref{fig:Epsilon2} also presents the sensitivity projection for an ideal experimental plan of one-year run at an event rate of 500 MHz with 100$\%$ duty factor.

\section{Summary}
\label{sec:summary}

A high-precision experimental setup is proposed at the Heavy Ion Accelerator Facility (HIAF) to explore potential new physics and test the SM through the decay channel $\eta \rightarrow e^{+}e^{-} \gamma$. The proposed one-month run time of the experiment is expected to yield approximately $5.9 \times 10^{11}$ $\eta$ mesons and a corresponding large data set of inelastic scattering events, generated through proton bombardment of a lithium target at 1.8 GeV. The $\eta$ meson production cross-section is modeled using the GiBUU event generator. To assess the performance of the conceptual spectrometer design and the physics potential of the proposed experiment, a detailed simulation framework has been developed. The signal events were simulated using custom-coded programs under two different detector setups, while GiBUU was employed to simulate the background events. The detector response and efficiency were modeled with the ChnsRoot package, which is based on the FairRoot framework.

Our study demonstrates that the designed spectrometer achieves a high detection efficiency and satisfactory energy and invariant mass resolutions for identifying the decay products. The upper limits on the branching ratio for the dark photon decay channel have been projected to be between $10^{-6}-10^{-7}$ at the $99\%$ confidence level in the range of $0 < m_A < 0.4$ GeV. This study highlights the capability of the proposed HIAF spectrometer for high-sensitivity searches and lays the foundation for future investigations into the properties of $\eta$-meson decay dynamics and new physics scenarios.

\section{Acknowledgement}
This work is supported by the CAS Project for Young Scientists in Basic Research YSBR-088. X. L. Kang acknowledges funding from the Guangdong Basic and Applied Basic Research Foundation Grant No. 2024A1515012416. We are grateful to our colleagues at the Institute of Modern Physics for their fruitful discussions.

\bibliographystyle{apsrev4-2}
\bibliography{ref.bib}

\begin{thebibliography}{47}%
\makeatletter
\providecommand \@ifxundefined [1]{%
 \@ifx{#1\undefined}
}%
\providecommand \@ifnum [1]{%
 \ifnum #1\expandafter \@firstoftwo
 \else \expandafter \@secondoftwo
 \fi
}%
\providecommand \@ifx [1]{%
 \ifx #1\expandafter \@firstoftwo
 \else \expandafter \@secondoftwo
 \fi
}%
\providecommand \natexlab [1]{#1}%
\providecommand \enquote  [1]{``#1''}%
\providecommand \bibnamefont  [1]{#1}%
\providecommand \bibfnamefont [1]{#1}%
\providecommand \citenamefont [1]{#1}%
\providecommand \href@noop [0]{\@secondoftwo}%
\providecommand \href [0]{\begingroup \@sanitize@url \@href}%
\providecommand \@href[1]{\@@startlink{#1}\@@href}%
\providecommand \@@href[1]{\endgroup#1\@@endlink}%
\providecommand \@sanitize@url [0]{\catcode `\\12\catcode `\$12\catcode
  `\&12\catcode `\#12\catcode `\^12\catcode `\_12\catcode `\%12\relax}%
\providecommand \@@startlink[1]{}%
\providecommand \@@endlink[0]{}%
\providecommand \url  [0]{\begingroup\@sanitize@url \@url }%
\providecommand \@url [1]{\endgroup\@href {#1}{\urlprefix }}%
\providecommand \urlprefix  [0]{URL }%
\providecommand \Eprint [0]{\href }%
\providecommand \doibase [0]{http://dx.doi.org/}%
\providecommand \selectlanguage [0]{\@gobble}%
\providecommand \bibinfo  [0]{\@secondoftwo}%
\providecommand \bibfield  [0]{\@secondoftwo}%
\providecommand \translation [1]{[#1]}%
\providecommand \BibitemOpen [0]{}%
\providecommand \bibitemStop [0]{}%
\providecommand \bibitemNoStop [0]{.\EOS\space}%
\providecommand \EOS [0]{\spacefactor3000\relax}%
\providecommand \BibitemShut  [1]{\csname bibitem#1\endcsname}%
\let\auto@bib@innerbib\@empty
\bibitem [{\citenamefont {Essig}\ \emph {et~al.}(2013)\citenamefont {Essig}
  \emph {et~al.}}]{Essig:2013lka}%
  \BibitemOpen
  \bibfield  {author} {\bibinfo {author} {\bibfnamefont {R.}~\bibnamefont
  {Essig}} \emph {et~al.},\ }in\ \href@noop {} {\emph {\bibinfo {booktitle}
  {{Snowmass 2013}: {Snowmass on the Mississippi}}}}\ (\bibinfo {year} {2013})\
  \Eprint {http://arxiv.org/abs/1311.0029} {arXiv:1311.0029 [hep-ph]}
  \BibitemShut {NoStop}%
\bibitem [{\citenamefont {Bertone}\ \emph {et~al.}(2005)\citenamefont
  {Bertone}, \citenamefont {Hooper},\ and\ \citenamefont
  {Silk}}]{Bertone:2004pz}%
  \BibitemOpen
  \bibfield  {author} {\bibinfo {author} {\bibfnamefont {G.}~\bibnamefont
  {Bertone}}, \bibinfo {author} {\bibfnamefont {D.}~\bibnamefont {Hooper}}, \
  and\ \bibinfo {author} {\bibfnamefont {J.}~\bibnamefont {Silk}},\ }\href
  {\doibase 10.1016/j.physrep.2004.08.031} {\bibfield  {journal} {\bibinfo
  {journal} {Phys. Rept.}\ }\textbf {\bibinfo {volume} {405}},\ \bibinfo
  {pages} {279} (\bibinfo {year} {2005})},\ \Eprint
  {http://arxiv.org/abs/hep-ph/0404175} {arXiv:hep-ph/0404175} \BibitemShut
  {NoStop}%
\bibitem [{\citenamefont {Beringer}\ \emph {et~al.}(2012)\citenamefont
  {Beringer} \emph {et~al.}}]{ParticleDataGroup:2012pjm}%
  \BibitemOpen
  \bibfield  {author} {\bibinfo {author} {\bibfnamefont {J.}~\bibnamefont
  {Beringer}} \emph {et~al.} (\bibinfo {collaboration} {Particle Data Group}),\
  }\href {\doibase 10.1103/PhysRevD.86.010001} {\bibfield  {journal} {\bibinfo
  {journal} {Phys. Rev. D}\ }\textbf {\bibinfo {volume} {86}},\ \bibinfo
  {pages} {010001} (\bibinfo {year} {2012})}\BibitemShut {NoStop}%
\bibitem [{\citenamefont {Krasznahorkay}\ \emph {et~al.}(2016)\citenamefont
  {Krasznahorkay} \emph {et~al.}}]{Krasznahorkay:2015iga}%
  \BibitemOpen
  \bibfield  {author} {\bibinfo {author} {\bibfnamefont {A.~J.}\ \bibnamefont
  {Krasznahorkay}} \emph {et~al.},\ }\href {\doibase
  10.1103/PhysRevLett.116.042501} {\bibfield  {journal} {\bibinfo  {journal}
  {Phys. Rev. Lett.}\ }\textbf {\bibinfo {volume} {116}},\ \bibinfo {pages}
  {042501} (\bibinfo {year} {2016})},\ \Eprint
  {http://arxiv.org/abs/1504.01527} {arXiv:1504.01527 [nucl-ex]} \BibitemShut
  {NoStop}%
\bibitem [{\citenamefont {Krasznahorkay}\ \emph {et~al.}(2019)\citenamefont
  {Krasznahorkay} \emph {et~al.}}]{Krasznahorkay:2019lyl}%
  \BibitemOpen
  \bibfield  {author} {\bibinfo {author} {\bibfnamefont {A.~J.}\ \bibnamefont
  {Krasznahorkay}} \emph {et~al.},\ }\href@noop {} {\  (\bibinfo {year}
  {2019})},\ \Eprint {http://arxiv.org/abs/1910.10459} {arXiv:1910.10459
  [nucl-ex]} \BibitemShut {NoStop}%
\bibitem [{\citenamefont {Pospelov}\ \emph {et~al.}(2008)\citenamefont
  {Pospelov}, \citenamefont {Ritz},\ and\ \citenamefont
  {Voloshin}}]{Pospelov:2007mp}%
  \BibitemOpen
  \bibfield  {author} {\bibinfo {author} {\bibfnamefont {M.}~\bibnamefont
  {Pospelov}}, \bibinfo {author} {\bibfnamefont {A.}~\bibnamefont {Ritz}}, \
  and\ \bibinfo {author} {\bibfnamefont {M.~B.}\ \bibnamefont {Voloshin}},\
  }\href {\doibase 10.1016/j.physletb.2008.02.052} {\bibfield  {journal}
  {\bibinfo  {journal} {Phys. Lett. B}\ }\textbf {\bibinfo {volume} {662}},\
  \bibinfo {pages} {53} (\bibinfo {year} {2008})},\ \Eprint
  {http://arxiv.org/abs/0711.4866} {arXiv:0711.4866 [hep-ph]} \BibitemShut
  {NoStop}%
\bibitem [{\citenamefont {Holdom}(1986)}]{Holdom:1985ag}%
  \BibitemOpen
  \bibfield  {author} {\bibinfo {author} {\bibfnamefont {B.}~\bibnamefont
  {Holdom}},\ }\href {\doibase 10.1016/0370-2693(86)91377-8} {\bibfield
  {journal} {\bibinfo  {journal} {Phys. Lett. B}\ }\textbf {\bibinfo {volume}
  {166}},\ \bibinfo {pages} {196} (\bibinfo {year} {1986})}\BibitemShut
  {NoStop}%
\bibitem [{\citenamefont {Fayet}(2007)}]{Fayet:2007ua}%
  \BibitemOpen
  \bibfield  {author} {\bibinfo {author} {\bibfnamefont {P.}~\bibnamefont
  {Fayet}},\ }\href {\doibase 10.1103/PhysRevD.75.115017} {\bibfield  {journal}
  {\bibinfo  {journal} {Phys. Rev. D}\ }\textbf {\bibinfo {volume} {75}},\
  \bibinfo {pages} {115017} (\bibinfo {year} {2007})},\ \Eprint
  {http://arxiv.org/abs/hep-ph/0702176} {arXiv:hep-ph/0702176} \BibitemShut
  {NoStop}%
\bibitem [{\citenamefont {Beranek}\ and\ \citenamefont
  {Vanderhaeghen}(2013)}]{Beranek:2012ey}%
  \BibitemOpen
  \bibfield  {author} {\bibinfo {author} {\bibfnamefont {T.}~\bibnamefont
  {Beranek}}\ and\ \bibinfo {author} {\bibfnamefont {M.}~\bibnamefont
  {Vanderhaeghen}},\ }\href {\doibase 10.1103/PhysRevD.87.015024} {\bibfield
  {journal} {\bibinfo  {journal} {Phys. Rev. D}\ }\textbf {\bibinfo {volume}
  {87}},\ \bibinfo {pages} {015024} (\bibinfo {year} {2013})},\ \Eprint
  {http://arxiv.org/abs/1209.4561} {arXiv:1209.4561 [hep-ph]} \BibitemShut
  {NoStop}%
\bibitem [{\citenamefont {Raggi}\ and\ \citenamefont
  {Kozhuharov}(2014)}]{Raggi:2014zpa}%
  \BibitemOpen
  \bibfield  {author} {\bibinfo {author} {\bibfnamefont {M.}~\bibnamefont
  {Raggi}}\ and\ \bibinfo {author} {\bibfnamefont {V.}~\bibnamefont
  {Kozhuharov}},\ }\href {\doibase 10.1155/2014/959802} {\bibfield  {journal}
  {\bibinfo  {journal} {Adv. High Energy Phys.}\ }\textbf {\bibinfo {volume}
  {2014}},\ \bibinfo {pages} {959802} (\bibinfo {year} {2014})},\ \Eprint
  {http://arxiv.org/abs/1403.3041} {arXiv:1403.3041 [physics.ins-det]}
  \BibitemShut {NoStop}%
\bibitem [{\citenamefont {Beranek}\ \emph {et~al.}(2013)\citenamefont
  {Beranek}, \citenamefont {Merkel},\ and\ \citenamefont
  {Vanderhaeghen}}]{Beranek:2013yqa}%
  \BibitemOpen
  \bibfield  {author} {\bibinfo {author} {\bibfnamefont {T.}~\bibnamefont
  {Beranek}}, \bibinfo {author} {\bibfnamefont {H.}~\bibnamefont {Merkel}}, \
  and\ \bibinfo {author} {\bibfnamefont {M.}~\bibnamefont {Vanderhaeghen}},\
  }\href {\doibase 10.1103/PhysRevD.88.015032} {\bibfield  {journal} {\bibinfo
  {journal} {Phys. Rev. D}\ }\textbf {\bibinfo {volume} {88}},\ \bibinfo
  {pages} {015032} (\bibinfo {year} {2013})},\ \Eprint
  {http://arxiv.org/abs/1303.2540} {arXiv:1303.2540 [hep-ph]} \BibitemShut
  {NoStop}%
\bibitem [{\citenamefont {Tulin}\ and\ \citenamefont
  {Yu}(2018)}]{Tulin:2017ara}%
  \BibitemOpen
  \bibfield  {author} {\bibinfo {author} {\bibfnamefont {S.}~\bibnamefont
  {Tulin}}\ and\ \bibinfo {author} {\bibfnamefont {H.-B.}\ \bibnamefont {Yu}},\
  }\href {\doibase 10.1016/j.physrep.2017.11.004} {\bibfield  {journal}
  {\bibinfo  {journal} {Phys. Rept.}\ }\textbf {\bibinfo {volume} {730}},\
  \bibinfo {pages} {1} (\bibinfo {year} {2018})},\ \Eprint
  {http://arxiv.org/abs/1705.02358} {arXiv:1705.02358 [hep-ph]} \BibitemShut
  {NoStop}%
\bibitem [{\citenamefont {Alexander}\ \emph {et~al.}(2016)\citenamefont
  {Alexander} \emph {et~al.}}]{Alexander:2016aln}%
  \BibitemOpen
  \bibfield  {author} {\bibinfo {author} {\bibfnamefont {J.}~\bibnamefont
  {Alexander}} \emph {et~al.}\ }(\bibinfo {year} {2016})\ \Eprint
  {http://arxiv.org/abs/1608.08632} {arXiv:1608.08632 [hep-ph]} \BibitemShut
  {NoStop}%
\bibitem [{\citenamefont {Battaglieri}\ \emph {et~al.}(2017)\citenamefont
  {Battaglieri} \emph {et~al.}}]{Battaglieri:2017aum}%
  \BibitemOpen
  \bibfield  {author} {\bibinfo {author} {\bibfnamefont {M.}~\bibnamefont
  {Battaglieri}} \emph {et~al.},\ }in\ \href@noop {} {\emph {\bibinfo
  {booktitle} {{U.S. Cosmic Visions: New Ideas in Dark Matter}}}}\ (\bibinfo
  {year} {2017})\ \Eprint {http://arxiv.org/abs/1707.04591} {arXiv:1707.04591
  [hep-ph]} \BibitemShut {NoStop}%
\bibitem [{\citenamefont {Arkani-Hamed}\ \emph {et~al.}(2009)\citenamefont
  {Arkani-Hamed}, \citenamefont {Finkbeiner}, \citenamefont {Slatyer},\ and\
  \citenamefont {Weiner}}]{Arkani-Hamed:2008hhe}%
  \BibitemOpen
  \bibfield  {author} {\bibinfo {author} {\bibfnamefont {N.}~\bibnamefont
  {Arkani-Hamed}}, \bibinfo {author} {\bibfnamefont {D.~P.}\ \bibnamefont
  {Finkbeiner}}, \bibinfo {author} {\bibfnamefont {T.~R.}\ \bibnamefont
  {Slatyer}}, \ and\ \bibinfo {author} {\bibfnamefont {N.}~\bibnamefont
  {Weiner}},\ }\href {\doibase 10.1103/PhysRevD.79.015014} {\bibfield
  {journal} {\bibinfo  {journal} {Phys. Rev. D}\ }\textbf {\bibinfo {volume}
  {79}},\ \bibinfo {pages} {015014} (\bibinfo {year} {2009})},\ \Eprint
  {http://arxiv.org/abs/0810.0713} {arXiv:0810.0713 [hep-ph]} \BibitemShut
  {NoStop}%
\bibitem [{\citenamefont {Bjorken}\ \emph {et~al.}(2009)\citenamefont
  {Bjorken}, \citenamefont {Essig}, \citenamefont {Schuster},\ and\
  \citenamefont {Toro}}]{Bjorken:2009mm}%
  \BibitemOpen
  \bibfield  {author} {\bibinfo {author} {\bibfnamefont {J.~D.}\ \bibnamefont
  {Bjorken}}, \bibinfo {author} {\bibfnamefont {R.}~\bibnamefont {Essig}},
  \bibinfo {author} {\bibfnamefont {P.}~\bibnamefont {Schuster}}, \ and\
  \bibinfo {author} {\bibfnamefont {N.}~\bibnamefont {Toro}},\ }\href {\doibase
  10.1103/PhysRevD.80.075018} {\bibfield  {journal} {\bibinfo  {journal} {Phys.
  Rev. D}\ }\textbf {\bibinfo {volume} {80}},\ \bibinfo {pages} {075018}
  (\bibinfo {year} {2009})},\ \Eprint {http://arxiv.org/abs/0906.0580}
  {arXiv:0906.0580 [hep-ph]} \BibitemShut {NoStop}%
\bibitem [{\citenamefont {Aaij}\ \emph {et~al.}(2020)\citenamefont {Aaij} \emph
  {et~al.}}]{LHCb:2019vmc}%
  \BibitemOpen
  \bibfield  {author} {\bibinfo {author} {\bibfnamefont {R.}~\bibnamefont
  {Aaij}} \emph {et~al.} (\bibinfo {collaboration} {LHCb}),\ }\href {\doibase
  10.1103/PhysRevLett.124.041801} {\bibfield  {journal} {\bibinfo  {journal}
  {Phys. Rev. Lett.}\ }\textbf {\bibinfo {volume} {124}},\ \bibinfo {pages}
  {041801} (\bibinfo {year} {2020})},\ \Eprint
  {http://arxiv.org/abs/1910.06926} {arXiv:1910.06926 [hep-ex]} \BibitemShut
  {NoStop}%
\bibitem [{\citenamefont {Graham}\ \emph {et~al.}(2021)\citenamefont {Graham},
  \citenamefont {Hearty},\ and\ \citenamefont {Williams}}]{Graham:2021ggy}%
  \BibitemOpen
  \bibfield  {author} {\bibinfo {author} {\bibfnamefont {M.}~\bibnamefont
  {Graham}}, \bibinfo {author} {\bibfnamefont {C.}~\bibnamefont {Hearty}}, \
  and\ \bibinfo {author} {\bibfnamefont {M.}~\bibnamefont {Williams}},\ }\href
  {\doibase 10.1146/annurev-nucl-110320-051823} {\bibfield  {journal} {\bibinfo
   {journal} {Ann. Rev. Nucl. Part. Sci.}\ }\textbf {\bibinfo {volume} {71}},\
  \bibinfo {pages} {37} (\bibinfo {year} {2021})},\ \Eprint
  {http://arxiv.org/abs/2104.10280} {arXiv:2104.10280 [hep-ph]} \BibitemShut
  {NoStop}%
\bibitem [{\citenamefont {Bross}\ \emph {et~al.}(1991)\citenamefont {Bross},
  \citenamefont {Crisler}, \citenamefont {Pordes}, \citenamefont {Volk},
  \citenamefont {Errede},\ and\ \citenamefont {Wrbanek}}]{Bross:1989mp}%
  \BibitemOpen
  \bibfield  {author} {\bibinfo {author} {\bibfnamefont {A.}~\bibnamefont
  {Bross}}, \bibinfo {author} {\bibfnamefont {M.}~\bibnamefont {Crisler}},
  \bibinfo {author} {\bibfnamefont {S.~H.}\ \bibnamefont {Pordes}}, \bibinfo
  {author} {\bibfnamefont {J.}~\bibnamefont {Volk}}, \bibinfo {author}
  {\bibfnamefont {S.}~\bibnamefont {Errede}}, \ and\ \bibinfo {author}
  {\bibfnamefont {J.}~\bibnamefont {Wrbanek}},\ }\href {\doibase
  10.1103/PhysRevLett.67.2942} {\bibfield  {journal} {\bibinfo  {journal}
  {Phys. Rev. Lett.}\ }\textbf {\bibinfo {volume} {67}},\ \bibinfo {pages}
  {2942} (\bibinfo {year} {1991})}\BibitemShut {NoStop}%
\bibitem [{\citenamefont {Riordan}\ \emph {et~al.}(1987)\citenamefont {Riordan}
  \emph {et~al.}}]{Riordan:1987aw}%
  \BibitemOpen
  \bibfield  {author} {\bibinfo {author} {\bibfnamefont {E.~M.}\ \bibnamefont
  {Riordan}} \emph {et~al.},\ }\href {\doibase 10.1103/PhysRevLett.59.755}
  {\bibfield  {journal} {\bibinfo  {journal} {Phys. Rev. Lett.}\ }\textbf
  {\bibinfo {volume} {59}},\ \bibinfo {pages} {755} (\bibinfo {year}
  {1987})}\BibitemShut {NoStop}%
\bibitem [{\citenamefont {Bl\"umlein}\ and\ \citenamefont
  {Brunner}(2014)}]{Blumlein:2013cua}%
  \BibitemOpen
  \bibfield  {author} {\bibinfo {author} {\bibfnamefont {J.}~\bibnamefont
  {Bl\"umlein}}\ and\ \bibinfo {author} {\bibfnamefont {J.}~\bibnamefont
  {Brunner}},\ }\href {\doibase 10.1016/j.physletb.2014.02.029} {\bibfield
  {journal} {\bibinfo  {journal} {Phys. Lett. B}\ }\textbf {\bibinfo {volume}
  {731}},\ \bibinfo {pages} {320} (\bibinfo {year} {2014})},\ \Eprint
  {http://arxiv.org/abs/1311.3870} {arXiv:1311.3870 [hep-ph]} \BibitemShut
  {NoStop}%
\bibitem [{\citenamefont {Gninenko}(2012)}]{Gninenko:2012eq}%
  \BibitemOpen
  \bibfield  {author} {\bibinfo {author} {\bibfnamefont {S.~N.}\ \bibnamefont
  {Gninenko}},\ }\href {\doibase 10.1016/j.physletb.2012.06.002} {\bibfield
  {journal} {\bibinfo  {journal} {Phys. Lett. B}\ }\textbf {\bibinfo {volume}
  {713}},\ \bibinfo {pages} {244} (\bibinfo {year} {2012})},\ \Eprint
  {http://arxiv.org/abs/1204.3583} {arXiv:1204.3583 [hep-ph]} \BibitemShut
  {NoStop}%
\bibitem [{\citenamefont {Merkel}\ \emph {et~al.}(2014)\citenamefont {Merkel}
  \emph {et~al.}}]{Merkel:2014avp}%
  \BibitemOpen
  \bibfield  {author} {\bibinfo {author} {\bibfnamefont {H.}~\bibnamefont
  {Merkel}} \emph {et~al.},\ }\href {\doibase 10.1103/PhysRevLett.112.221802}
  {\bibfield  {journal} {\bibinfo  {journal} {Phys. Rev. Lett.}\ }\textbf
  {\bibinfo {volume} {112}},\ \bibinfo {pages} {221802} (\bibinfo {year}
  {2014})},\ \Eprint {http://arxiv.org/abs/1404.5502} {arXiv:1404.5502
  [hep-ex]} \BibitemShut {NoStop}%
\bibitem [{\citenamefont {Batley}\ \emph {et~al.}(2015)\citenamefont {Batley}
  \emph {et~al.}}]{NA482:2015wmo}%
  \BibitemOpen
  \bibfield  {author} {\bibinfo {author} {\bibfnamefont {J.~R.}\ \bibnamefont
  {Batley}} \emph {et~al.} (\bibinfo {collaboration} {NA48/2}),\ }\href
  {\doibase 10.1016/j.physletb.2015.04.068} {\bibfield  {journal} {\bibinfo
  {journal} {Phys. Lett. B}\ }\textbf {\bibinfo {volume} {746}},\ \bibinfo
  {pages} {178} (\bibinfo {year} {2015})},\ \Eprint
  {http://arxiv.org/abs/1504.00607} {arXiv:1504.00607 [hep-ex]} \BibitemShut
  {NoStop}%
\bibitem [{\citenamefont {Archilli}\ \emph {et~al.}(2012)\citenamefont
  {Archilli} \emph {et~al.}}]{KLOE-2:2011hhj}%
  \BibitemOpen
  \bibfield  {author} {\bibinfo {author} {\bibfnamefont {F.}~\bibnamefont
  {Archilli}} \emph {et~al.} (\bibinfo {collaboration} {KLOE-2}),\ }\href
  {\doibase 10.1016/j.physletb.2011.11.033} {\bibfield  {journal} {\bibinfo
  {journal} {Phys. Lett. B}\ }\textbf {\bibinfo {volume} {706}},\ \bibinfo
  {pages} {251} (\bibinfo {year} {2012})},\ \Eprint
  {http://arxiv.org/abs/1110.0411} {arXiv:1110.0411 [hep-ex]} \BibitemShut
  {NoStop}%
\bibitem [{\citenamefont {Babusci}\ \emph {et~al.}(2013)\citenamefont {Babusci}
  \emph {et~al.}}]{KLOE-2:2012lii}%
  \BibitemOpen
  \bibfield  {author} {\bibinfo {author} {\bibfnamefont {D.}~\bibnamefont
  {Babusci}} \emph {et~al.} (\bibinfo {collaboration} {KLOE-2}),\ }\href
  {\doibase 10.1016/j.physletb.2013.01.067} {\bibfield  {journal} {\bibinfo
  {journal} {Phys. Lett. B}\ }\textbf {\bibinfo {volume} {720}},\ \bibinfo
  {pages} {111} (\bibinfo {year} {2013})},\ \Eprint
  {http://arxiv.org/abs/1210.3927} {arXiv:1210.3927 [hep-ex]} \BibitemShut
  {NoStop}%
\bibitem [{\citenamefont {Adlarson}\ \emph {et~al.}(2013)\citenamefont
  {Adlarson} \emph {et~al.}}]{WASA-at-COSY:2013zom}%
  \BibitemOpen
  \bibfield  {author} {\bibinfo {author} {\bibfnamefont {P.}~\bibnamefont
  {Adlarson}} \emph {et~al.} (\bibinfo {collaboration} {WASA-at-COSY}),\ }\href
  {\doibase 10.1016/j.physletb.2013.08.055} {\bibfield  {journal} {\bibinfo
  {journal} {Phys. Lett. B}\ }\textbf {\bibinfo {volume} {726}},\ \bibinfo
  {pages} {187} (\bibinfo {year} {2013})},\ \Eprint
  {http://arxiv.org/abs/1304.0671} {arXiv:1304.0671 [hep-ex]} \BibitemShut
  {NoStop}%
\bibitem [{\citenamefont {Agakishiev}\ \emph {et~al.}(2014)\citenamefont
  {Agakishiev} \emph {et~al.}}]{HADES:2013nab}%
  \BibitemOpen
  \bibfield  {author} {\bibinfo {author} {\bibfnamefont {G.}~\bibnamefont
  {Agakishiev}} \emph {et~al.} (\bibinfo {collaboration} {HADES}),\ }\href
  {\doibase 10.1016/j.physletb.2014.02.035} {\bibfield  {journal} {\bibinfo
  {journal} {Phys. Lett. B}\ }\textbf {\bibinfo {volume} {731}},\ \bibinfo
  {pages} {265} (\bibinfo {year} {2014})},\ \Eprint
  {http://arxiv.org/abs/1311.0216} {arXiv:1311.0216 [hep-ex]} \BibitemShut
  {NoStop}%
\bibitem [{\citenamefont {Adare}\ \emph {et~al.}(2015)\citenamefont {Adare}
  \emph {et~al.}}]{PHENIX:2014duq}%
  \BibitemOpen
  \bibfield  {author} {\bibinfo {author} {\bibfnamefont {A.}~\bibnamefont
  {Adare}} \emph {et~al.} (\bibinfo {collaboration} {PHENIX}),\ }\href
  {\doibase 10.1103/PhysRevC.91.031901} {\bibfield  {journal} {\bibinfo
  {journal} {Phys. Rev. C}\ }\textbf {\bibinfo {volume} {91}},\ \bibinfo
  {pages} {031901} (\bibinfo {year} {2015})},\ \Eprint
  {http://arxiv.org/abs/1409.0851} {arXiv:1409.0851 [nucl-ex]} \BibitemShut
  {NoStop}%
\bibitem [{\citenamefont {Anastasi}\ \emph {et~al.}(2016)\citenamefont
  {Anastasi} \emph {et~al.}}]{KLOE-2:2016ydq}%
  \BibitemOpen
  \bibfield  {author} {\bibinfo {author} {\bibfnamefont {A.}~\bibnamefont
  {Anastasi}} \emph {et~al.} (\bibinfo {collaboration} {KLOE-2}),\ }\href
  {\doibase 10.1016/j.physletb.2016.04.019} {\bibfield  {journal} {\bibinfo
  {journal} {Phys. Lett. B}\ }\textbf {\bibinfo {volume} {757}},\ \bibinfo
  {pages} {356} (\bibinfo {year} {2016})},\ \Eprint
  {http://arxiv.org/abs/1603.06086} {arXiv:1603.06086 [hep-ex]} \BibitemShut
  {NoStop}%
\bibitem [{\citenamefont {Chen}\ \emph {et~al.}(2024)\citenamefont {Chen} \emph
  {et~al.}}]{Chen:2024wad}%
  \BibitemOpen
  \bibfield  {author} {\bibinfo {author} {\bibfnamefont {X.-R.}\ \bibnamefont
  {Chen}} \emph {et~al.},\ }\href@noop {} {\  (\bibinfo {year} {2024})},\
  \Eprint {http://arxiv.org/abs/2407.00874} {arXiv:2407.00874 [hep-ph]}
  \BibitemShut {NoStop}%
\bibitem [{\citenamefont {Ruegg}\ and\ \citenamefont
  {Ruiz-Altaba}(2004)}]{Ruegg:2003ps}%
  \BibitemOpen
  \bibfield  {author} {\bibinfo {author} {\bibfnamefont {H.}~\bibnamefont
  {Ruegg}}\ and\ \bibinfo {author} {\bibfnamefont {M.}~\bibnamefont
  {Ruiz-Altaba}},\ }\href {\doibase 10.1142/S0217751X04019755} {\bibfield
  {journal} {\bibinfo  {journal} {Int. J. Mod. Phys. A}\ }\textbf {\bibinfo
  {volume} {19}},\ \bibinfo {pages} {3265} (\bibinfo {year} {2004})},\ \Eprint
  {http://arxiv.org/abs/hep-th/0304245} {arXiv:hep-th/0304245} \BibitemShut
  {NoStop}%
\bibitem [{\citenamefont {Lanfranchi}\ \emph {et~al.}(2021)\citenamefont
  {Lanfranchi}, \citenamefont {Pospelov},\ and\ \citenamefont
  {Schuster}}]{Lanfranchi:2020crw}%
  \BibitemOpen
  \bibfield  {author} {\bibinfo {author} {\bibfnamefont {G.}~\bibnamefont
  {Lanfranchi}}, \bibinfo {author} {\bibfnamefont {M.}~\bibnamefont
  {Pospelov}}, \ and\ \bibinfo {author} {\bibfnamefont {P.}~\bibnamefont
  {Schuster}},\ }\href {\doibase 10.1146/annurev-nucl-102419-055056} {\bibfield
   {journal} {\bibinfo  {journal} {Ann. Rev. Nucl. Part. Sci.}\ }\textbf
  {\bibinfo {volume} {71}},\ \bibinfo {pages} {279} (\bibinfo {year} {2021})},\
  \Eprint {http://arxiv.org/abs/2011.02157} {arXiv:2011.02157 [hep-ph]}
  \BibitemShut {NoStop}%
\bibitem [{\citenamefont {Patrignani}\ \emph {et~al.}(2016)\citenamefont
  {Patrignani} \emph {et~al.}}]{ParticleDataGroup:2016lqr}%
  \BibitemOpen
  \bibfield  {author} {\bibinfo {author} {\bibfnamefont {C.}~\bibnamefont
  {Patrignani}} \emph {et~al.} (\bibinfo {collaboration} {Particle Data
  Group}),\ }\href {\doibase 10.1088/1674-1137/40/10/100001} {\bibfield
  {journal} {\bibinfo  {journal} {Chin. Phys. C}\ }\textbf {\bibinfo {volume}
  {40}},\ \bibinfo {pages} {100001} (\bibinfo {year} {2016})}\BibitemShut
  {NoStop}%
\bibitem [{\citenamefont {Lees}\ \emph {et~al.}(2014)\citenamefont {Lees} \emph
  {et~al.}}]{BaBar:2014zli}%
  \BibitemOpen
  \bibfield  {author} {\bibinfo {author} {\bibfnamefont {J.~P.}\ \bibnamefont
  {Lees}} \emph {et~al.} (\bibinfo {collaboration} {BaBar}),\ }\href {\doibase
  10.1103/PhysRevLett.113.201801} {\bibfield  {journal} {\bibinfo  {journal}
  {Phys. Rev. Lett.}\ }\textbf {\bibinfo {volume} {113}},\ \bibinfo {pages}
  {201801} (\bibinfo {year} {2014})},\ \Eprint {http://arxiv.org/abs/1406.2980}
  {arXiv:1406.2980 [hep-ex]} \BibitemShut {NoStop}%
\bibitem [{\citenamefont {Izaguirre}\ \emph {et~al.}(2015)\citenamefont
  {Izaguirre}, \citenamefont {Krnjaic}, \citenamefont {Schuster},\ and\
  \citenamefont {Toro}}]{Izaguirre:2015yja}%
  \BibitemOpen
  \bibfield  {author} {\bibinfo {author} {\bibfnamefont {E.}~\bibnamefont
  {Izaguirre}}, \bibinfo {author} {\bibfnamefont {G.}~\bibnamefont {Krnjaic}},
  \bibinfo {author} {\bibfnamefont {P.}~\bibnamefont {Schuster}}, \ and\
  \bibinfo {author} {\bibfnamefont {N.}~\bibnamefont {Toro}},\ }\href {\doibase
  10.1103/PhysRevLett.115.251301} {\bibfield  {journal} {\bibinfo  {journal}
  {Phys. Rev. Lett.}\ }\textbf {\bibinfo {volume} {115}},\ \bibinfo {pages}
  {251301} (\bibinfo {year} {2015})},\ \Eprint
  {http://arxiv.org/abs/1505.00011} {arXiv:1505.00011 [hep-ph]} \BibitemShut
  {NoStop}%
\bibitem [{\citenamefont {Fayet}(2004)}]{Fayet:2004bw}%
  \BibitemOpen
  \bibfield  {author} {\bibinfo {author} {\bibfnamefont {P.}~\bibnamefont
  {Fayet}},\ }\href {\doibase 10.1103/PhysRevD.70.023514} {\bibfield  {journal}
  {\bibinfo  {journal} {Phys. Rev. D}\ }\textbf {\bibinfo {volume} {70}},\
  \bibinfo {pages} {023514} (\bibinfo {year} {2004})},\ \Eprint
  {http://arxiv.org/abs/hep-ph/0403226} {arXiv:hep-ph/0403226} \BibitemShut
  {NoStop}%
\bibitem [{\citenamefont {Adrian}\ \emph {et~al.}(2018)\citenamefont {Adrian}
  \emph {et~al.}}]{HPS:2018xkw}%
  \BibitemOpen
  \bibfield  {author} {\bibinfo {author} {\bibfnamefont {P.~H.}\ \bibnamefont
  {Adrian}} \emph {et~al.} (\bibinfo {collaboration} {HPS}),\ }\href {\doibase
  10.1103/PhysRevD.98.091101} {\bibfield  {journal} {\bibinfo  {journal} {Phys.
  Rev. D}\ }\textbf {\bibinfo {volume} {98}},\ \bibinfo {pages} {091101}
  (\bibinfo {year} {2018})},\ \Eprint {http://arxiv.org/abs/1807.11530}
  {arXiv:1807.11530 [hep-ex]} \BibitemShut {NoStop}%
\bibitem [{\citenamefont {Ablikim}\ \emph {et~al.}(2024)\citenamefont {Ablikim}
  \emph {et~al.}}]{BESIII:2024pxo}%
  \BibitemOpen
  \bibfield  {author} {\bibinfo {author} {\bibfnamefont {M.}~\bibnamefont
  {Ablikim}} \emph {et~al.} (\bibinfo {collaboration} {BESIII}),\ }\href
  {\doibase 10.1103/PhysRevD.109.072001} {\bibfield  {journal} {\bibinfo
  {journal} {Phys. Rev. D}\ }\textbf {\bibinfo {volume} {109}},\ \bibinfo
  {pages} {072001} (\bibinfo {year} {2024})},\ \Eprint
  {http://arxiv.org/abs/2401.09136} {arXiv:2401.09136 [hep-ex]} \BibitemShut
  {NoStop}%
\bibitem [{\citenamefont {Raggi}\ \emph {et~al.}(2015)\citenamefont {Raggi},
  \citenamefont {Kozhuharov},\ and\ \citenamefont {Valente}}]{Raggi:2015gza}%
  \BibitemOpen
  \bibfield  {author} {\bibinfo {author} {\bibfnamefont {M.}~\bibnamefont
  {Raggi}}, \bibinfo {author} {\bibfnamefont {V.}~\bibnamefont {Kozhuharov}}, \
  and\ \bibinfo {author} {\bibfnamefont {P.}~\bibnamefont {Valente}},\ }\href
  {\doibase 10.1051/epjconf/20159601025} {\bibfield  {journal} {\bibinfo
  {journal} {EPJ Web Conf.}\ }\textbf {\bibinfo {volume} {96}},\ \bibinfo
  {pages} {01025} (\bibinfo {year} {2015})},\ \Eprint
  {http://arxiv.org/abs/1501.01867} {arXiv:1501.01867 [hep-ex]} \BibitemShut
  {NoStop}%
\bibitem [{\citenamefont {Elam}\ \emph {et~al.}(2022)\citenamefont {Elam} \emph
  {et~al.}}]{REDTOP:2022slw}%
  \BibitemOpen
  \bibfield  {author} {\bibinfo {author} {\bibfnamefont {J.}~\bibnamefont
  {Elam}} \emph {et~al.} (\bibinfo {collaboration} {REDTOP}),\ }\href@noop {}
  {\  (\bibinfo {year} {2022})},\ \Eprint {http://arxiv.org/abs/2203.07651}
  {arXiv:2203.07651 [hep-ex]} \BibitemShut {NoStop}%
\bibitem [{\citenamefont {An}\ \emph {et~al.}(2025)\citenamefont {An} \emph
  {et~al.}}]{An:2025lws}%
  \BibitemOpen
  \bibfield  {author} {\bibinfo {author} {\bibfnamefont {F.}~\bibnamefont {An}}
  \emph {et~al.},\ }\href@noop {} {\  (\bibinfo {year} {2025})},\ \Eprint
  {http://arxiv.org/abs/2504.21050} {arXiv:2504.21050 [hep-ph]} \BibitemShut
  {NoStop}%
\bibitem [{\citenamefont {Gatto}\ \emph {et~al.}(2022)\citenamefont {Gatto},
  \citenamefont {Blazey}, \citenamefont {Dychkant}, \citenamefont {Elam},
  \citenamefont {Figora}, \citenamefont {Fletcher}, \citenamefont {Francis},
  \citenamefont {Liu}, \citenamefont {Los}, \citenamefont {Mahieu},
  \citenamefont {Mane}, \citenamefont {Marquez}, \citenamefont {Murray},
  \citenamefont {Ramberg}, \citenamefont {Royon}, \citenamefont {Syphers},
  \citenamefont {Young},\ and\ \citenamefont {Zutshi}}]{instruments6040049}%
  \BibitemOpen
  \bibfield  {author} {\bibinfo {author} {\bibfnamefont {C.}~\bibnamefont
  {Gatto}}, \bibinfo {author} {\bibfnamefont {G.~C.}\ \bibnamefont {Blazey}},
  \bibinfo {author} {\bibfnamefont {A.}~\bibnamefont {Dychkant}}, \bibinfo
  {author} {\bibfnamefont {J.~W.}\ \bibnamefont {Elam}}, \bibinfo {author}
  {\bibfnamefont {M.}~\bibnamefont {Figora}}, \bibinfo {author} {\bibfnamefont
  {T.}~\bibnamefont {Fletcher}}, \bibinfo {author} {\bibfnamefont
  {K.}~\bibnamefont {Francis}}, \bibinfo {author} {\bibfnamefont
  {A.}~\bibnamefont {Liu}}, \bibinfo {author} {\bibfnamefont {S.}~\bibnamefont
  {Los}}, \bibinfo {author} {\bibfnamefont {C.~L.}\ \bibnamefont {Mahieu}},
  \bibinfo {author} {\bibfnamefont {A.~U.}\ \bibnamefont {Mane}}, \bibinfo
  {author} {\bibfnamefont {J.}~\bibnamefont {Marquez}}, \bibinfo {author}
  {\bibfnamefont {M.~J.}\ \bibnamefont {Murray}}, \bibinfo {author}
  {\bibfnamefont {E.}~\bibnamefont {Ramberg}}, \bibinfo {author} {\bibfnamefont
  {C.}~\bibnamefont {Royon}}, \bibinfo {author} {\bibfnamefont {M.~J.}\
  \bibnamefont {Syphers}}, \bibinfo {author} {\bibfnamefont {R.~W.}\
  \bibnamefont {Young}}, \ and\ \bibinfo {author} {\bibfnamefont
  {V.}~\bibnamefont {Zutshi}},\ }\href {\doibase 10.3390/instruments6040049}
  {\bibfield  {journal} {\bibinfo  {journal} {Instruments}\ }\textbf {\bibinfo
  {volume} {6}} (\bibinfo {year} {2022}),\
  10.3390/instruments6040049}\BibitemShut {NoStop}%
\bibitem [{\citenamefont {Buss}\ \emph {et~al.}(2012)\citenamefont {Buss},
  \citenamefont {Gaitanos}, \citenamefont {Gallmeister}, \citenamefont {van
  Hees}, \citenamefont {Kaskulov}, \citenamefont {Lalakulich}, \citenamefont
  {Larionov}, \citenamefont {Leitner}, \citenamefont {Weil},\ and\
  \citenamefont {Mosel}}]{Buss:2011mx}%
  \BibitemOpen
  \bibfield  {author} {\bibinfo {author} {\bibfnamefont {O.}~\bibnamefont
  {Buss}}, \bibinfo {author} {\bibfnamefont {T.}~\bibnamefont {Gaitanos}},
  \bibinfo {author} {\bibfnamefont {K.}~\bibnamefont {Gallmeister}}, \bibinfo
  {author} {\bibfnamefont {H.}~\bibnamefont {van Hees}}, \bibinfo {author}
  {\bibfnamefont {M.}~\bibnamefont {Kaskulov}}, \bibinfo {author}
  {\bibfnamefont {O.}~\bibnamefont {Lalakulich}}, \bibinfo {author}
  {\bibfnamefont {A.~B.}\ \bibnamefont {Larionov}}, \bibinfo {author}
  {\bibfnamefont {T.}~\bibnamefont {Leitner}}, \bibinfo {author} {\bibfnamefont
  {J.}~\bibnamefont {Weil}}, \ and\ \bibinfo {author} {\bibfnamefont
  {U.}~\bibnamefont {Mosel}},\ }\href {\doibase 10.1016/j.physrep.2011.12.001}
  {\bibfield  {journal} {\bibinfo  {journal} {Phys. Rept.}\ }\textbf {\bibinfo
  {volume} {512}},\ \bibinfo {pages} {1} (\bibinfo {year} {2012})},\ \Eprint
  {http://arxiv.org/abs/1106.1344} {arXiv:1106.1344 [hep-ph]} \BibitemShut
  {NoStop}%
\bibitem [{\citenamefont {Gaitanos}\ \emph {et~al.}(2008)\citenamefont
  {Gaitanos}, \citenamefont {Lenske},\ and\ \citenamefont
  {Mosel}}]{Gaitanos:2007mm}%
  \BibitemOpen
  \bibfield  {author} {\bibinfo {author} {\bibfnamefont {T.}~\bibnamefont
  {Gaitanos}}, \bibinfo {author} {\bibfnamefont {H.}~\bibnamefont {Lenske}}, \
  and\ \bibinfo {author} {\bibfnamefont {U.}~\bibnamefont {Mosel}},\ }\href
  {\doibase 10.1016/j.physletb.2008.04.011} {\bibfield  {journal} {\bibinfo
  {journal} {Phys. Lett. B}\ }\textbf {\bibinfo {volume} {663}},\ \bibinfo
  {pages} {197} (\bibinfo {year} {2008})},\ \Eprint
  {http://arxiv.org/abs/0712.3292} {arXiv:0712.3292 [nucl-th]} \BibitemShut
  {NoStop}%
\bibitem [{\citenamefont {Weil}\ \emph {et~al.}(2012)\citenamefont {Weil},
  \citenamefont {van Hees},\ and\ \citenamefont {Mosel}}]{Weil:2012ji}%
  \BibitemOpen
  \bibfield  {author} {\bibinfo {author} {\bibfnamefont {J.}~\bibnamefont
  {Weil}}, \bibinfo {author} {\bibfnamefont {H.}~\bibnamefont {van Hees}}, \
  and\ \bibinfo {author} {\bibfnamefont {U.}~\bibnamefont {Mosel}},\ }\href
  {\doibase 10.1140/epja/i2012-12111-9} {\bibfield  {journal} {\bibinfo
  {journal} {Eur. Phys. J. A}\ }\textbf {\bibinfo {volume} {48}},\ \bibinfo
  {pages} {111} (\bibinfo {year} {2012})},\ \bibinfo {note} {[Erratum:
  Eur.Phys.J.A 48, 150 (2012)]},\ \Eprint {http://arxiv.org/abs/1203.3557}
  {arXiv:1203.3557 [nucl-th]} \BibitemShut {NoStop}%
\bibitem [{\citenamefont {Aguar-Bartolome}\ \emph {et~al.}(2014)\citenamefont
  {Aguar-Bartolome} \emph {et~al.}}]{A2:2013wad}%
  \BibitemOpen
  \bibfield  {author} {\bibinfo {author} {\bibfnamefont {P.}~\bibnamefont
  {Aguar-Bartolome}} \emph {et~al.} (\bibinfo {collaboration} {A2}),\ }\href
  {\doibase 10.1103/PhysRevC.89.044608} {\bibfield  {journal} {\bibinfo
  {journal} {Phys. Rev. C}\ }\textbf {\bibinfo {volume} {89}},\ \bibinfo
  {pages} {044608} (\bibinfo {year} {2014})},\ \Eprint
  {http://arxiv.org/abs/1309.5648} {arXiv:1309.5648 [hep-ex]} \BibitemShut
  {NoStop}%
\end{thebibliography}%

\end{document}